\documentclass[11pt]{JHEP3}

\JHEPspecialurl{http://jhep.sissa.it/JOURNAL/JHEP3.tar.gz}

\usepackage{epsfig,multicol}
\usepackage{amsmath}

\newcommand\fverb{\setbox\fverbbox=\hbox\bgroup\verb}
\newcommand\fverbdo{\egroup\medskip\noindent%
			\fbox{\unhbox\fverbbox}\ }
\newcommand\fverbit{\egroup\item[\fbox{\unhbox\fverbbox}]}
\newbox\fverbbox

\newcommand{\nn}{\nonumber}
\def\dfrac#1#2{\displaystyle\frac{#1}{#2}}

\newcommand{\pslash}{p\kern-1ex /}
\newcommand{\qslash}{q\kern-1ex /}
\newcommand{\lslash}{l\kern-1ex /}
\newcommand{\sslash}{s\kern-1ex /}
\newcommand{\kaslash}{k_a\kern-2ex /}
\newcommand{\kbslash}{k_b\kern-2ex /}
\newcommand{\Dslash}{\mathcal{D}\kern-1.5ex /}

\newcommand{\tr}{\mathrm{tr}}

\newcommand{\beqa}{\begin{eqnarray}}
\newcommand{\eeqa}{\end{eqnarray}}
\newcommand{\rmO}{\mathrm{O}}
\newcommand{\rmd}{\mathrm{d}}
\newcommand{\rme}{\mathrm{e}}
\newcommand{\ba}{\begin{eqnarray}}
\newcommand{\ea}{\end{eqnarray}}
\newcommand{\be}{\begin{equation}}   
\newcommand{\Nf}{N_\mathrm{f}}

\title{Application of the operator product expansion to the short distance
behavior of nuclear potentials}
\author{Sinya Aoki\\
       Graduate School of Pure and Applied Sciences, University of Tsukuba, Tsukuba, Ibaraki 305-8571,    Japan\\
        E-mail: \email{saoki@het.ph.tsukuba.ac.jp}}

\author{Janos Balog\\
        Research Institute for Particle and Nuclear Physics, 1523 Budapest 114, Pf. 49, Hungary\\
        E-mail: \email{balog@rmki.kfki.hu}}
 \author{Peter Weisz\\
        Max-Planck-Institut f\"ur Physik, F\"ohringer Ring 6, D-80805 M\"unchen, Germany\\
        E-mail: \email{pew@mpp.mpg.de}}       

\received{\today} 		
\accepted{\today}	
\preprint{MPP-2010-14, UTHEP-603}

\abstract{We investigate the short distance behavior of 
nucleon--nucleon (NN) potentials defined through Bethe--Salpeter wave 
functions, by perturbatively calculating anomalous dimensions of 6--quark 
operators in QCD. Thanks to the asymptotic freedom of QCD, 
1--loop computations give certain exact results 
for the potentials in the zero distance limit. In particular
the functional form of the S--state central NN 
potential at short distance $r$ is predicted to be a little weaker 
than $r^{-2}$. On the other hand, due to the  
intriguing character of the anomalous dimension spectrum, 
perturbative considerations alone can not determine whether this potential 
is repulsive or attractive at short distances. 
A crude estimation suggests that the force at short distance is repulsive, 
as found numerically in lattice QCD.
A similar behavior is found for the tensor potential.}


\begin{document}

\section{Introduction}

In a recent paper \cite{IAH} a proposal has been made 
to study nucleon--nucleon (NN) potentials from a first principle 
QCD approach. In this field theoretic framework, potentials are 
obtained through the Schr\"odinger operator applied to
Bethe--Salpeter (BS) wave functions defined by
\beqa
\varphi_E(\vec r) &=& \langle 0 \vert N(\vec x+\vec r, t)N(\vec x , t) 
\vert 2{\rm N}, E\rangle\,,
\eeqa
where $\vert 2{\rm N}, E\rangle$ is a QCD eigenstate with energy $E$ 
(suppressing here other quantum numbers),
and $N$ is a nucleon interpolating operator made of 3 quarks such as 
$N(x) =\epsilon^{abc}q^a(x)q^b(x)q^c(x)$.
Such wave functions have been measured through numerical simulations
of the lattice regularized theory \cite{IAH,AHI1,AHI2,IAH2}. 
Although many conceptual questions remain to be resolved,
the corresponding potentials 
indeed qualitatively resemble phenomenological NN potentials which 
are widely used in nuclear physics. The force at medium to long distance 
($r\ge 2$ fm) is shown to be attractive. This feature has long well been 
understood in terms of pion and other heavier meson exchanges.
At short distance, a characteristic repulsive core is reproduced by the 
lattice QCD simulation \cite{IAH}. No simple theoretical explanation, 
however, exists so far for the origin of the repulsive core. 
For an approach based on string theories, see ref. \cite{string}.

By writing
\beqa
\langle 0 \vert  N(\vec x+\vec r, t) N(\vec x , t)  
&=& \sum_{n=0}^\infty \int \frac{\rmd E}{2E}  \langle 2{\rm N}, n\pi, E\vert 
f_n(\vec r, E)\,,
\eeqa
where $\vert 2{\rm N}, n\pi, E\rangle$ is a state with the energy $E$
containing two nucleons and $n$ pions (and/or nucleon-antinucleon pairs),
we see that $\varphi_E(\vec r) = f_0(\vec r, E)$. 
(Our normalization is 
$\langle 2{\rm N}, n\pi, E \vert 2{\rm N}, n'\pi, E'\rangle
=2E\delta_{nn'}\delta(E-E')$.) 
We may thus interpret the wave function $\varphi_E(\vec r)$ 
as an amplitude to find the QCD eigenstate $\vert 2{\rm N}, E\rangle$ in
$N(\vec x+\vec r, t) N(\vec x , t) \vert 0 \rangle$.

The behavior of the wave functions $\varphi_E(\vec r)$ 
at short distances ($r=\vert \vec r\vert$) 
are encoded in the operator product expansion (OPE) of 
$ N(\vec x+\vec r, t) N(\vec x , t)$. 
An OPE analysis \cite{ABW1} of BS wave functions in the case of a toy model, 
the Ising field theory in 2--dimensions,
successfully described the analytically known behavior.
In this case the limiting short distance behavior of the potential 
does not depend on the energy (rapidity) of the state, 
and further it only mildly depends on energy (for low energies) 
at distances of the order of the Compton wave length of the particles.

In this report we perform an operator product expansion (OPE) analysis
of NN BS wave functions in QCD, with the aim to theoretically 
better understand the repulsive core of the NN potential, 
(at least that of the measured BS potential).  
Thanks to the property of asymptotic freedom of QCD
the form of leading short distance behavior of the coefficent functions can
be computed using perturbation theory. A short summary of our results has 
been published in ref. \cite{ABW2}\footnote{Unfortunately the results for
$\beta^{01}$ and $\beta^{10}$ as given in ref. \cite{ABW2} differ 
(incorrectly) from (\ref{summary2}) by a factor of 2.}.

In sect.~2 we start with some general considerations on BS potentials,
and sect.~3 presents some standard renormalization group 
considerations. The anomalous dimensions of 3-- and 6--quark operators
are computed in sect.~4. Finally in sects.~5,~6 we discuss the application 
of the results to NN potentials. 
In appendix~C we make a similar analysis for the $I=2$ two pion system.
For the convenience of the reader we give a 
brief summary of our results here.
The OPE analysis shows that the NN central potential at short distance behaves 
as
\beqa
V_c^{SI} (r) &\simeq& C_E \frac{(-\log r )^{\beta^{SI} -1}}{r^2}
\label{summary1}
\eeqa
for the S--state ($L=0$) , where $S$ and $I$ are total spin and isospin of 
the NN system, respectively, and $\beta^{SI}$ is negative 
and explicitly obtained as
\beqa
\beta^{01} &=& - \frac{6}{33-2\Nf}, \quad
\beta^{10} = - \frac{2}{33-2\Nf}\,,
\label{summary2}
\eeqa
where $\Nf$ is the number of quark flavors, and the overall coefficient 
$C_E$  depends on the energy $E$. 
Unfortunately the OPE analysis is not as conclusive 
as that in the toy model referred to above, in particular
the sign of $C_E$ is not determined by perturbative cosiderations alone.
The latter requires additional non-perturbative knowledge of matrix 
elements of composite operators.
A crude estimation using the non-relativistic quark model 
indicates that $C_E$ is positive, which implies a repulsive core
with a potential diverging a little weaker than the 
generically expected $r^{-2}$ at short distances.

\section{Operator Product Expansion and potentials at short distance in 3 dimensions}
\label{sec:3dim}

In this section we discuss the application of the operator product expansion
(OPE) to the determination of the short distance behavior of the BS potential.
We consider the equal time Bethe--Salpeter (BS) wave function defined by
\begin{eqnarray}
\varphi_{AB}^E(\vec r) &=& 
\langle 0 \vert O_A(\vec r/2, 0) O_B(-\vec r/2, 0) \vert E \rangle\,,
\end{eqnarray}
where $\vert E \rangle$ is an eigen-state of the system with energy $E$, 
and  $O_A$, $O_B$ are some operators of the system. 
Here we suppress other quantum numbers of the state $\vert E\rangle $ 
for simplicity.
Using the OPE of $O_A$ and $O_B$ 
\begin{eqnarray}
O_A (\vec r/2, 0) O_B(-\vec r/2,0) &\simeq& 
\sum_C D_{AB}^C(\vec r) O_C(\vec 0, 0)\,, 
\label{ope22}
\end{eqnarray}
we have
\begin{eqnarray}
\varphi_{AB}^E(\vec r) &\simeq& 
\sum_C D_{AB}^C(\vec r) \langle 0 \vert O_C(\vec 0, 0)\vert E \rangle\,.
\end{eqnarray}
Note that $\vec r$ dependence appears solely in $D_{AB}^C(\vec r)$ 
while the $E$ dependence exists only in 
$\langle 0 \vert O_C(\vec 0,0) \vert E \rangle$.   
As we will see, in the $r=\vert \vec r\vert \rightarrow 0$ limit, 
the coefficient function behaves as
\begin{eqnarray}
D_{AB}^C (\vec r) \simeq  r^{\alpha_C} (-\log r)^{\beta_C} f_C(\theta,\phi)\,, 
\label{alphabeta}
\end{eqnarray}
where $\theta,\phi$ are angles in the polar coordinates of $\vec r$,
so that
\begin{eqnarray}
\varphi_{AB}^E (\vec r) \simeq \sum_C r^{\alpha_C}(-\log r)^{\beta_C} 
f_C(\theta,\phi)D_C(E)\,, \quad
D_C(E) = \langle 0 \vert O_C (\vec 0,0)\vert E \rangle\,.
\end{eqnarray}
We now assume that $C$ has the largest contribution at small $r$:
\begin{eqnarray}
\alpha_C &<&  \alpha_{C'} \quad \mbox{or} \\
\alpha_C &=& \alpha_{C'}, \quad \beta_C > \beta_{C'}\,.
\end{eqnarray}
for $^\forall C'\not= C$.
The potential can be calculated from this wave function. 

As will be seen later, $\alpha_C=\alpha_{C^\prime} =0$ for the NN case in QCD. 
Furthermore states with zero orbital angular momentum ($L=0$) dominates in the OPE, so that
the wave function at short distance is given by
\begin{eqnarray}
\varphi_{AB}^E(r) \simeq \left[  (-\log r)^{\beta_C} D_C(E)  + 
(-\log r)^{\beta_{C^\prime}} D_{C^\prime}(E) \right] 
\end{eqnarray}
with $\beta_C > \beta_{C^\prime}$. Using
\begin{eqnarray}
\nabla^2  (-\log r)^{\beta}
&=& -\beta (-\log r)^{\beta-1}
\left[ 1- \frac{\beta-1}{-\log r}\right] r^{-2} \,,
\end{eqnarray}
we obtain the following classification of the short distance behavior
of the potential. 
\begin{enumerate}
\item $\beta_C \not=0$: The potential at short distance is energy independent and becomes
\begin{eqnarray} V(r) &\simeq& -\frac{\beta_C}{r^2(-\log r)}\,,
\end{eqnarray}
which is attractive for $\beta_C >0 $ and repulsive for $\beta_C < 0$.
\item $\beta_C =0$: In this case we have
\begin{eqnarray}
V(r) &\simeq& \frac{D_{C'}(E)}{D_C(E)}
\left(\frac{-\beta_{C'}}{r^2}\right)
(-\log r)^{\beta_{C^\prime}-1}\,.
\end{eqnarray}
The sign of the potential at short distance depends on 
$ -\beta_{C^\prime}D_{C'}(E)  / D_C(E)$. 
\end{enumerate}
If there are two or more operators which have the largest contribution
at short distance, we have
\begin{eqnarray}
\varphi_{AB}^E(x) &=& (-\log r)^{\beta_C}( D_{C_1}(E) + D_{C_2}(E)+\cdots)\,.
\end{eqnarray}
In this case, the above analysis can be applied just by replacing $D_C(E)
\rightarrow D_{C_1}(E) + D_{C_2}(E)+\cdots$.

On the lattice, we do not expect divergences at $r=0$ due to 
lattice artifacts at short distance.
The above classification holds at $a \ll r \ll 1\,{\rm fm}$, 
while the potential becomes finite even at $r=0$ on the lattice.

\section{Renormalization group analysis and operator product expansion}

\subsection{Renormalization group equation for composite operators}

In QCD, using dimensional regularization in $D=4-2\epsilon$ dimensions,
bare local composite operators $O^{(0)}_A(x)$ are renormalized \cite{RG}
according to\footnote{We note that we are considering
the massless theory here since quark masses play no role in our analysis.}
\begin{equation}
O^{({\rm ren})}_A(x)=Z_{AB}(g,\epsilon)\,O^{(0)}_B(x).
\end{equation}
(Summation of repeated indices is assumed throughout this paper.)
The meaning of the above formula is that we obtain finite results
if we insert the right hand side into any correlation function, provided
we also renormalize the bare QCD coupling $g_0$ and the quark and gluon fields.
For example, in the case of an $n$--quark correlation function with operator
insertion, which we denote by ${\cal G}^{(0)}_{n;A}(g_0,\epsilon)$ 
(suppressing the dependence on the quark momenta and other quantum numbers)
we have
\begin{equation}
{\cal G}^{({\rm ren})}_{n;A}(g,\mu)=Z_{AB}(g,\epsilon)\,
Z_F^{-n/2}(g,\epsilon)\, 
{\cal G}^{(0)}_{n;B}(g_0,\epsilon). 
\end{equation}
We recall from renormalization theory
that for the analogous $n$--quark 
correlation function (without any operator insertion) we have
\begin{equation}
{\cal G}^{({\rm ren})}_n(g,\mu)=Z_F^{-n/2}(g,\epsilon)\, 
{\cal G}^{(0)}_n(g_0,\epsilon), 
\end{equation}
where the coupling renormalization is given by
\begin{equation}
g_0^2=\mu^{2\epsilon}\,Z_1(g,\epsilon)\,g^2.
\end{equation}
The renormalization constant $Z_1$ in the minimal subtraction (MS)
scheme we are using has pure pole terms only:
\begin{equation}
Z_1(g,\epsilon)=1-\frac{\beta_0g^2}{\epsilon}-\frac{\beta_1g^4}{2\epsilon}
+\frac{\beta_0^2g^4}{\epsilon^2}+\rmO(g^6),
\end{equation}
where
\begin{equation}
\beta_0=\frac{1}{16\pi^2}\left\{\frac{11}{3}N-\frac{2}{3}N_f\right\},
\qquad\quad
\beta_1=\frac{1}{256\pi^4}\left\{\frac{34}{3}N^2-\left(
\frac{13}{3}N-\frac{1}{N}\right)N_f\right\}.
\end{equation}
Similarly for the fermion field renormalization constant, we have
\begin{equation}
Z_F(g,\epsilon)=1-\frac{\gamma_{F0}g^2}{2\epsilon}+\rmO(g^4),
\end{equation}
where $\gamma_{F0}$ is given by (\ref{wfrc}).
The gluon field renormalization constant is also similar, but we do not need 
it here. Finally the matrix of operator renormalization constants
is of the form
\begin{equation}
Z_{AB}(g,\epsilon)=\delta_{AB}-\frac{\gamma^{(1)}_{AB}g^2}{2\epsilon}
+\rmO(g^4).
\end{equation}

The renormalization group (RG) expresses the simple fact that bare quantities
are independent of the renormalization scale $\mu$. Introducing the RG 
differential operator
\begin{equation}
{\cal D}=\mu\frac{\partial}{\partial\mu}+\beta(g)\,\frac{\partial}{\partial g}
\end{equation}
the RG equation for $n$--quark correlation functions can be written as
\begin{equation}
\left\{{\cal D}+\frac{n}{2}\gamma_F(g)\right\}\,
{\cal G}^{({\rm ren})}_n(g,\mu)=0,
\end{equation}
where the RG beta function is
\begin{equation}
\beta(g)=\epsilon g +\beta_D(g,\epsilon)
=\epsilon g-\frac{\epsilon g}{1+\frac{g}{2}\,
\frac{\partial\ln Z_1}{\partial g}}=-\beta_0g^3-\beta_1g^5+\rmO(g^7),
\end{equation}
where $\beta_D(g,\epsilon)$ is the beta function in $D$ dimenions
and the RG gamma function (for quark fields) is
\begin{equation}
\gamma_F(g)=\beta_D(g,\epsilon)\,\frac{\partial\ln Z_F}
{\partial g}=\gamma_{F0}\,g^2+\rmO(g^4).
\end{equation}
It is useful to introduce the RG invariant lambda-parameter $\Lambda$
by taking the Ansatz
\begin{equation}
\Lambda=\mu\,{\rm e}^{f(g)}
\end{equation}
and requiring ${\cal D}\Lambda=0$. The solution is the lambda-parameter
in the MS scheme ($\Lambda_{\rm MS}$) if the arbitrary integration constant
is fixed by requiring that for small coupling
\begin{equation}
f(g)=-\frac{1}{2\beta_0g^2}-\frac{\beta_1}{2\beta_0^2}\,\ln(\beta_0g^2)
+\rmO(g^2).
\end{equation}
Finally the RG equations for $n$--quark correlation functions with
operator insertion are of the form
\begin{equation}
\left\{{\cal D}+\frac{n}{2}\gamma_F(g)\right\}\,
{\cal G}^{({\rm ren})}_{n;A}(g,\mu)-\gamma_{AB}(g)
{\cal G}^{({\rm ren})}_{n;B}(g,\mu)=0,
\end{equation}
where
\begin{equation}
\gamma_{AB}(g)=-Z_{AC}\beta_D(g,\epsilon)\frac
{\partial Z^{-1}_{CB}}{\partial g}=\gamma^{(1)}_{AB}g^2+\rmO(g^4).
\end{equation}

\subsection{OPE and RG equations}

Let us recall the operator product expansion (\ref{ope22})
\begin{equation}
O_1(y/2)O_2(-y/2)\simeq D_B(y)\,O_B(0).
\label{ope33}
\end{equation}
We will need it in the special case where the operators $O_1,O_2$ on the
left hand side are nucleon operators and the set of operators $O_B$ on
the right hand side are local 6--quark operators of engineering dimension 9
and higher. All operators in (\ref{ope33}) are renormalized ones, but from 
now on we suppress the labels $^{({\rm ren})}$.
As we will see, the nucleon operators are renormalized diagonally as
\begin{equation}
O_1=\zeta_1(g,\epsilon)\,O^{(0)}_1,\qquad\qquad
O_2=\zeta_2(g,\epsilon)\,O^{(0)}_2,
\end{equation}
and we can define the corresponding RG gamma functions by
\begin{equation}
\gamma_{1,2}(g)=\beta_D(g,\epsilon)\,
\frac{\partial\ln\zeta_{1,2}}{\partial g}=\gamma_{1,2}^{(1)}g^2+\rmO(g^4).
\end{equation}

Next we write down the bare version of (\ref{ope33}) (in terms of bare 
operators and bare coefficient functions):
\begin{equation}
O^{(0)}_1(y/2)O_2^{(0)}(-y/2)\simeq D^{(0)}_B(y)\,O^{(0)}_B(0).
\label{ope33bare}
\end{equation}
Comparing (\ref{ope33}) to (\ref{ope33bare}), we can read off the 
renormalization of the coefficient functions:
\begin{equation}
D_B(y)=\zeta_1(g,\epsilon)\zeta_2(g,\epsilon)D^{(0)}_A(y)\,
Z^{-1}_{AB}(g,\epsilon)
\end{equation}
and using the $\mu$-independence of the bare coefficient functions
we can derive the RG equations satisfied by the renormalized ones:
\begin{equation}
{\cal D}D_B(g,\mu,y)+D_A(g,\mu,y)\,\tilde\gamma_{AB}(g)=0,
\label{RG33}
\end{equation}
where the effective gamma function matrix is defined as
\begin{equation}
\tilde\gamma_{AB}(g)=\gamma_{AB}(g)-\left[\gamma_1(g)+\gamma_2(g)\right]
\,\delta_{AB}.
\end{equation}

\subsection{Perturbative solution of the RG equation and factorization
of OPE}

We want to solve the vector partial differential equation (\ref{RG33}) and
for this purpose it is useful to introduce $\hat U_{AB}(g)$,
the solution of the matrix ordinary differential equation
\begin{equation}
\beta(g)\,\frac{{\rm d}}{{\rm d}g}\,\hat U_{AB}(g)=\tilde\gamma_{AC}(g)
\,\hat U_{CB}(g)
\label{hatU}
\end{equation}
and its matrix inverse $U_{AB}(g)$. We will assume that the coefficient
functions are dimensionless and have the perturbative expansion
\begin{equation}
D_A(g,\mu,y)=D_A(g;\mu r)=D_{A;0}+g^2D_{A;1}(\mu r)+\rmO(g^4),
\label{Dpert}
\end{equation}
where $r=\vert y\vert$.
For the case of operators with higher engineering dimension $9+\alpha$
the coefficients are of the form $r^\alpha$ times dimensionless
functions and the analysis is completely analogous and can be done
independently, since in the massless theory operators of different dimension
do not mix.
(In the full theory quark mass terms are also present, but they
correspond to higher powers in $r$ and therefore can be neglected.)

We will also assume that the basis of operators has been chosen such that the
1-loop mixing matrix is diagonal:
\begin{equation}
\tilde\gamma_{AB}(g)=2\beta_0\,\beta_A\,g^2\,\delta_{AB}+\rmO(g^4).
\end{equation}
In such a basis the solution of (\ref{hatU}) in perturbation theory
takes the form
\begin{equation}
\hat U_{AB}(g)=\left\{\delta_{AB}+R_{AB}(g)\right\}\,g^{-2\beta_B},
\label{hatUpert}
\end{equation}
where $R_{AB}(g)=\rmO(g^2)$, with possible multiplicative $\log g^2$ factors,
depending on the details of the spectrum of 1-loop eigenvalues $\beta_A$.

Having solved (\ref{hatU}) we can write down the most general solution
of (\ref{RG33}):
\begin{equation}
D_B(g;\mu r)=F_A(\Lambda r)\,U_{AB}(g).
\end{equation}
Here the vector $F_A$ is RG-invariant. Introducing the running
coupling $\bar g$ as the solution of the equation
\begin{equation}
f(\bar g)=f(g)+\ln(\mu r)=\ln(\Lambda r)
\end{equation}
$F_B$ can be rewritten as
\begin{equation}
F_B(\Lambda r)=D_A(\bar g;1)\,\hat U_{AB}(\bar g).
\end{equation}
Since, because of asymptotic freedom (AF), for $r\to0$ also $\bar g\to0$ as
\begin{equation}
\bar g^2\approx-\frac{1}{2\beta_0\ln(\Lambda r)},
\end{equation}
$F_B$ can be calculated perturbatively using (\ref{Dpert}) and 
(\ref{hatUpert}).

Putting everything together, we find that the right hand side of the
operator product expansion (\ref{ope33}) can be rewritten:
\begin{equation}
O_1(y/2)O_2(-y/2)\simeq F_B(\Lambda r)\,\tilde O_B(0),
\label{ope34}
\end{equation}
where
\begin{equation}
\tilde O_B=U_{BC}(g)\,O_C.
\end{equation}
There is a factorization of the operator product into perturbative and 
non-perturbative quantities: $F_B(\Lambda r)$ is perturbative and 
calculable (for $r\to0$) thanks to AF, whereas the matrix elements of
$\tilde O_B$ are non-perturbative (but $r$-independent).

An operator $O_B$ first occurring at $\ell_B$-loop order on the right hand 
side of (\ref{ope33}) and corresponding to normalized 1-loop eigenvalue
$\beta_B$ has coefficient $F_B(\Lambda r)$ with leading short distance
behavior
\begin{equation}
F_B(\Lambda r)\approx \bar g^{2(\ell_B-\beta_B)}\approx
\left(-2\beta_0\ln(\Lambda r)\right)^{\beta_B-\ell_B}.
\end{equation}
In principle, an operator with very large $\beta_B$, even if it is not
present in the expansion at tree level yet, might be important at short 
distances. This is why it is necessary to calculate the full 1-loop 
spectrum of $\beta_B$ eigenvalues.
As we shall see, no such operators exist in our cases, 
and therefore operators with
non-vanishing tree level coefficients are dominating at short distances.
The corresponding coefficient functions have leading short distance
behavior given by
\begin{equation}
F_B(\Lambda r)\approx D_{B;0}\,\left(-2\beta_0\ln(\Lambda r)\right)^{\beta_B}.
\end{equation}
A similar analysis in the case of operators of dimension $9+\alpha$
leads to the result (\ref{alphabeta}).


\section{OPE and Anomalous dimensions for two nucleons}


\subsection{OPE of two nucleon operators at tree level}
The general form of a gauge invariant local 3--quark operator is given by
\beqa
B^F_\Gamma (x) \equiv B^{fgh}_{\alpha\beta\gamma}(x) 
= \varepsilon^{abc} q^{a,f}_\alpha(x) q^{b,g}_\beta(x) q^{c,h}_\gamma (x)\,,
\label{baryonop}
\eeqa
where $\alpha,\beta,\gamma$ are spinor, $f,g,h$ are flavor, 
$a,b,c$ are color indices of the (renormalized) quark field $q$.  
The color index runs from 1 to $N=3$, the spinor index from 1 to 4, 
and the flavor index from 1 to $\Nf$. 
In this paper a summation over a repeated index is assumed, unless otherwise
stated. Note that $B^{fgh}_{\alpha\beta\gamma}$ is symmetric under any
interchange of pairs of indices 
(e.g. $B^{fgh}_{\alpha\beta\gamma}=B^{gfh}_{\beta\alpha\gamma}$)
because the quark fields anticommute.
For simplicity we sometimes use the 
notation such as $F=fgh$ and $\Gamma=\alpha\beta\gamma$ as indicated in
(\ref{baryonop}).

The nucleon operator is constructed from the above operators as
\beqa
B^f_\alpha(x) = \left(P_{+4}\right)_{\alpha\alpha'} 
B_{\alpha'\beta\gamma}^{fgh} (C\gamma_5)_{\beta\gamma}(i\tau_2)^{gh}\,,
\eeqa 
where $P_{+4} = (1+\gamma_4)/2$ is the projection to the large 
spinor component, $C=\gamma_2\gamma_4$ is the charge conjugation matrix, 
and $\tau_2$ is the Pauli matrix in the flavor space (for $\Nf=2$) given by 
$(i\tau_2)^{fg} = \varepsilon^{fg}$. 
Both $C\gamma_5$ and $i\tau_2$ are anti-symmetric under
the interchange of two indices, so that the nucleon operator 
has spin $1/2$ and isospin $1/2$.  Although the explicit form 
of the $\gamma$ matrices is unnecessary in principle, 
we find it convenient to use a (chiral) convention given by
\beqa
 \gamma_k &=& \left( \begin{array}{cc}
 0 & i\sigma_k \\
 -i\sigma_k & 0
 \end{array}
 \right)\,, \
 \gamma_4 = \left( \begin{array}{cc}
 0 & {\bf 1} \\
 {\bf 1} & 0 \\
 \end{array}
 \right)\,, \
 \gamma_5 =\gamma_1\gamma_2\gamma_3\gamma_4 =
 \left( \begin{array}{cc}
 {\bf 1}& 0  \\
 0 & -{\bf 1} \\
 \end{array}
 \right)\,.
 \eeqa

As discussed in the previous section, the OPE at the 
tree level (generically) dominates at short distance.
The OPE of two nucleon operators given above at tree level becomes 
\beqa
B^f_{\alpha}(x+y/2) B^g_{\beta}(x-y/2) &=& B^f_{\alpha}(x) B^g_{\beta}(x) 
+\frac{y^\mu}{2}\left\{ \partial_\mu[ B^f_\alpha (x)] B^g_\beta(x) 
- B^f_\alpha (x)\partial_\mu[ B^g_\beta(x)]
\right\}\nn \\
&+&\frac{y^\mu y^\nu}{8} \left\{\partial_\mu\partial_\nu 
[B^f_\alpha(x) B^g_\beta(x) ] 
-4\partial_\mu B^f_\alpha(x)\partial_\nu B^g_\beta(x) \right\}
+\cdots .
\eeqa
For the two-nucleon operator with either the combination 
$[\alpha\beta]$, $\{fg\}$  ($S=0$) 
or the combination $\{\alpha\beta\}$, $[fg]$ ($S=1$), 
terms odd in $y$ vanish in the above OPE, so that only even $L$ contributions 
appear. These 6--quark operators are anti-symmetric under the exchange 
$(\alpha,f)\leftrightarrow (\beta,g)$. 
On the other hand, for two other operators with $( [\alpha\beta], [fg])$ 
or $(\{\alpha\beta\},\{fg\})$, which are symmetric under the exchange,
terms even in $y$ vanish in the OPE and only odd $L$'s contribute.

Knowing the anomalous dimensions of the 6--quark operators appearing in the OPE, 
which will be calculated later in this section, 
the OPE at short distance ($r=\vert \vec y\vert \ll 1$, $y_4=0$)  becomes
\beqa
B^f_{\alpha}(x+y/2) B^g_{\beta}(x-y/2) &\simeq& \sum_A c_A(r) O^{fg,A}_{\alpha\beta}(x)
+\sum_B d_B(r) y^k y^l O^{fg,B}_{\alpha\beta,kl} (x) \nn \\ 
&+&\sum_C e_C(r) y^k O^{fg,C}_{\alpha\beta,k}(x) 
+\cdots \,,
\eeqa
where the coefficient functions behave as
\beqa
c_A (r) &\simeq& (-\log r)^{\beta_A}\,, \quad
d_B (r) \simeq   (-\log r)^{\beta_B}\,, \quad
e_C (r) \simeq   (-\log r)^{\beta_C}\,,
\eeqa
and $\beta_{A,B,C}$ are related to the anomalous dimensions of the 6--quark 
operators $O^{fg,A}_{\alpha\beta}$, of those with two derivatives 
$O^{fg,B}_{\alpha\beta,kl}$ and of those with one derivative 
$O^{fg,C}_{\alpha\beta,k}$.

The wave function defined through the eigenstate $\vert E\rangle$  is given by
\beqa
\varphi_E^{\rm even}(y) &=& \langle 0 \vert B^f_{\alpha}(x+y/2) 
B^g_{\beta}(x-y/2)\vert E\rangle \nn\\
&\simeq&
\sum_A c_A (r) \langle 0 \vert O^{fg,A}_{\alpha\beta}(x) \vert E\rangle
+ \sum_B d_B (r) y^k y^l \langle 0 \vert O^{fg,B}_{\alpha\beta,kl} (x) 
\vert E\rangle+\rmO(y^4)\,,
\eeqa
for the anti-symmetric states,
while 
\beqa
\varphi_E^{\rm odd}(y) &=& \langle 0 \vert B^f_{\alpha}(x+y/2) B^g_{\beta}(x-y/2)\vert E\rangle
\simeq \sum_C e_C (r) y^k  \langle 0 \vert O^{fg,C}_{\alpha\beta,k} (x) 
\vert E\rangle+\rmO(y^3)
\eeqa
for the symmetric states.

In this paper, we consider only 6--quark operators without derivatives  and calculate the corresponding anomalous dimensions. 

\subsection{General formula for the divergent  part at 1-loop}

Following the previous section, we define the renormalization factor
$Z_X$ of a $k$--quark operator $X=[q^k]$ through the relation 
\beqa
[q^k]^{\mathrm{ren}} &=& Z_X [q_0^k] = Z_X Z_F^{k/2}[q^k]\,,
\eeqa
where $q_0$($q$) is the bare (renormalized) quark field.
The wave function renormalization factor for the quark field 
is given at 1-loop by
\beqa
Z_F &=& 1 + g^2 Z_F^{(1)} \,, \quad Z_F^{(1)} 
= -\frac{\lambda C_F }{16\pi^2\epsilon}
\label{wfrc}
\eeqa
where $\lambda$ is the gauge parameter and $C_F =\frac{N^2-1}{2N}$.

At 1-loop the renormalization of simple $k$--quark operators 
(those involving no gauge fields) is given by the divergent
parts of diagrams involving exchange of a gluon between any
pair of quark fields.
The 1-loop correction to the insertion of an operator 
$q^{a,f}_\alpha(x) q^{b,g}_\beta (x)$ in any correlation function 
involving external quarks is expressed as the contraction of
\beqa
q^{a,f}_\alpha(x) q^{b,g}_\beta(x)  
\frac{1}{2!} \int \rmd^D y\, \rmd^D z \, A_\mu^A(y) A_\nu^B(z) 
[\bar q^{f_1}(y) ig T^A \gamma_\mu q^{f_1}(y)]
[\bar q^{g_1}(z) ig T^B \gamma_\nu q^{g_1}(z)]
\label{eq:op_1loop}
\eeqa
where $\tr\, T^A T^B =\delta_{AB}/2$ in our normalization.
Since two identical contributions cancel the $2!$ in the denominator,
the contraction at 1-loop is given by
\beqa
-g^2 (T^A)_{aa_1}(T^A)_{bb_1}\int \rmd^Dy\, \rmd^D z\, &&
\left[S_F(x-y)\gamma_\mu q(y)\right]_{\alpha}^{a_1f_1} G_{\mu\nu}(y-z)\nn \\
&\times& 
\left[S_F(x-z)\gamma_\nu q(z)\right]_{\beta}^{b_1g_1} 
\eeqa
where the free quark and gauge propagators are given in momentum space as
\beqa
S_F(p) &=& \frac{ - i \pslash + m}{p^2+ m^2}\,, \quad
G_{\mu\nu}(k) = \frac{1}{k^2}
\left[ g_{\mu\nu} - (1-\lambda)\frac{k_\mu k_\nu}{k^2}\right]\,.
\eeqa 
The above contribution can be written as
\beqa
\frac{g^2}{2N}\{\delta_{aa_1}\delta_{bb_1}-N\delta_{ab_1}\delta_{a_1b}\}
\int \frac{\rmd^D p\, \rmd^D q}{(2\pi)^{2D}} 
T_{\alpha\alpha_1,\beta\beta_1}(p,q)\, q^{a_1f_1}_{\alpha_1}(p)\rme^{ipx}
q^{b_1g_1}_{\beta_1}(q)\rme^{iqx}
\label{eq:contraction}
\eeqa
where
\beqa
T_{\alpha\alpha_1,\beta\beta_1}(p,q) &=&
\int \frac{\rmd^D k}{(2\pi)^D} \left[S_F(p+k) 
\gamma_\mu\right]_{\alpha\alpha_1}
G_{\mu\nu}(k) \left[S_F(q-k) \gamma_\nu\right]_{\beta\beta_1}\,,
\eeqa
whose divergent part is independent of the momenta $p,q$ and is given by
\beqa
T_{\alpha\alpha_1,\beta\beta_1}(0,0)
&=& \frac{1}{16\pi^2}\frac{1}{\epsilon} 
\left[ -\frac{1}{4} \sum_{\mu\nu} 
\sigma_{\mu\nu} \otimes \sigma_{\mu\nu} 
+ \lambda 1\otimes1\right]_{\alpha\alpha_1,\beta\beta_1} 
\eeqa
with $\sigma_{\mu\nu} =\frac{i}{2}\left[\gamma_\mu,\gamma_\nu\right]$.
We then obtain the divergent part of the 1-loop contribution as
\beqa
\left[ q^{a,f}_\alpha(x) q^{b,g}_\beta (x)\right]^{\rm 1-loop, div}
&=& \frac{g^2}{32 N\pi^2}\frac{1}{\epsilon} 
\left[ ({\bf T}_0 + \lambda {\bf T}_1) \cdot 
q^a(x)\otimes q^b (x)\right]_{\alpha,\beta}^{fg}
\label{eq:1-loopT}
\eeqa
where (bold--faced symbols represent matrices in flavor and spinor space) 
\beqa
({\bf T}_0)^{f f_1,g g_1}_{\alpha\alpha_1,\beta\beta_1} &=&
-\frac{1}{4}\sum_{\mu\nu}\left\{ 
\boldsymbol{\sigma}_{\mu\nu}\otimes\boldsymbol{\sigma}_{\mu\nu}
+ N \boldsymbol{\sigma}_{\mu\nu}\tilde\otimes\boldsymbol{\sigma}_{\mu\nu}
\right\}^{f f_1,g g_1}_{\alpha\alpha_1,\beta\beta_1}\,, \\
({\bf T}_1)^{f f_1,g g_1}_{\alpha\alpha_1,\beta\beta_1} &=&
\left\{{\bf 1}\otimes{\bf 1}
+ N{\bf 1}\tilde\otimes{\bf 1}\right\}^{f f_1,g 
g_1}_{\alpha\alpha_1,\beta\beta_1}\,.
\label{eq:T1}
\eeqa
Here we use the notation
\beqa
\{{\bf X}\otimes{\bf Y}\}^{f f_1,g g_1}_{\alpha\alpha_1,\beta\beta_1} &=&
{\bf X}^{f f_1}_{\alpha\alpha_1}{\bf Y}^{g g_1}_{\beta\beta_1} \qquad
\{{\bf X}\tilde\otimes{\bf Y}\}^{f f_1,g g_1}_{\alpha\alpha_1,\beta\beta_1} =
{\bf X}^{g f_1}_{\beta\alpha_1} {\bf Y}^{f g_1}_{\alpha\beta_1}\,, \\
\{\boldsymbol{\sigma}_{\mu\nu}\}^{f g}_{\alpha\beta}&=&\delta^{fg} 
(\sigma_{\mu\nu})_{\alpha\beta},
\quad \{{\bf 1}\}^{f g}_{\alpha\beta} = \delta^{fg}\delta_{\alpha\beta}\,.
\eeqa
Using the following Fierz identities for spinor indices
\beqa
-\frac{1}{4}\sum_{\mu\nu} \sigma_{\mu\nu}\otimes\sigma_{\mu\nu}
&=& P_R\otimes P_R + P_L\otimes P_L 
- 2( P_R\tilde\otimes P_R + P_L\tilde\otimes P_L)\,,
\label{eq:fierz1}\\
-\frac{1}{4}\sum_{\mu\nu} \sigma_{\mu\nu}\tilde\otimes 
\sigma_{\mu\nu}
&=& P_R\tilde\otimes P_R + P_L\tilde\otimes P_L 
- 2( P_R\otimes P_R + P_L\otimes P_L)\,,
\label{eq:fierz2}
\eeqa
where $P_R,P_L$ are the chiral projectors
\beqa
P_R&=&\frac12(1+\gamma_5)\,,\,\,\,\,\, P_L =\frac12(1-\gamma_5)\,,
\eeqa
we can simplify ${\bf T}_0$ as
\beqa
({\bf T}_0)^{f f_1,g g_1}_{\alpha\alpha_1,\beta\beta_1} &=& 
\delta^{ff_1}\delta^{gg_1}
\left[ \delta_{\alpha\alpha_1}\delta_{\beta\beta_1} 
    - 2\delta_{\beta\alpha_1}\delta_{\alpha\beta_1}\right]
+N\delta^{gf_1}\delta^{fg_1}
\left[ \delta_{\beta\alpha_1}\delta_{\alpha\beta_1}
-2\delta_{\alpha\alpha_1}\delta_{\beta\beta_1} \right]\nn \\
\label{eq:T0}
\eeqa
where either $\alpha_1,\beta_1 \in \{1,2\}$(right-handed) or 
$\alpha_1,\beta_1 \in \{3,4\}$(left-handed)
due to the chiral projections in eqs. (\ref{eq:fierz1}) 
and (\ref{eq:fierz2}). 
In our following calculation of the 1-loop anomalous dimensions, 
eq.~(\ref{eq:1-loopT}) together with eqs.~(\ref{eq:T0}) and (\ref{eq:T1})
are the key equations. 

\subsection{Renormalization of local 3--quark operators at 1-loop}

In this subsection we calculate the anomalous dimensions of 
general 3--quark operators at 1-loop. 
In terms of the renormalization factor defined as
\beqa
B_3^{\rm renor.} &=& \zeta [q_0^3] = \zeta Z_F^{3/2}[q^3] , \quad 
\zeta =  1 + g^2 (\zeta^{(1)} +\zeta_\lambda^{(1)} )+\dots\,,
\eeqa
where $\zeta^{(1)}$ ( $\zeta_\lambda^{(1)}$ ) is the $\lambda$--independent 
(dependent) part at 1-loop,
the divergent part of the insertion of the 3--quark operator  
$B^F_\Gamma=B_{\alpha\beta\gamma}^{fgh}$ defined in (\ref{baryonop})
at 1-loop is given by a linear combination of insertion of baryon operators,
and (with a slight abuse of notation) we express this as
\beqa
(\Gamma^{(1)\mathrm{div}})^F_\Gamma &=& - g^2 
\left(\zeta^{(1)}+\zeta_\lambda^{(1)}+\frac{3}{2}Z_F^{(1)}\right)
_{\Gamma\Gamma'}^{FF'}B_{\Gamma'}^{F'}\,.
\eeqa

The $\lambda$--dependent contribution from ${\bf T}_1$ in (\ref{eq:T1})
is diagonal and given by
\beqa
g^2(\Gamma^{(1)\mathrm{div}}_\lambda)_\Gamma^F &=& 
3\lambda \frac{g^2}{32 \pi^2}\frac{N+1}{N\epsilon} B_\Gamma^F\,,
\eeqa
so that the $\lambda$--dependent part of $\zeta$ vanishes:
\beqa
\zeta_\lambda^{(1)} &=&-\frac{3\lambda}{32 \pi^2}\frac{N+1}{N\epsilon} 
-\frac{3}{2} Z_F^{(1)}
=\frac{\lambda}{64 N\pi^2}\frac{3(N+1)(N-3)}{\epsilon} = 0\,, \quad (N=3)\,.
\eeqa
Therefore $\zeta$ is $\lambda$--independent, 
as expected from the gauge invariance.
We remark that we leave $N$ explicit in some formulae to help keep track
of the origin of the various terms, but in our case we should always set 
$N=3$ at the end.

The $\lambda$--independent part of $\Gamma^{(1)}$ from ${\bf T}_0$ 
in (\ref{eq:T0}) leads to $(N=3)$:
\beqa
(\Gamma^{(1)\mathrm{div}})^{fgh}_{\alpha\beta\gamma} 
&=&\frac{(N+1)}{2N}\frac{g^2}{16\pi^2\epsilon}
\left[3 B^{fgh}_{\alpha\beta\gamma} 
-2B^{fgh}_{\beta\alpha\gamma}
-2B^{fgh}_{\gamma\beta\alpha}
-2B^{fgh}_{\alpha\gamma\beta}\right]\,,
\\
(\Gamma^{(1)\mathrm{div}})^{fgh}_{\alpha\beta\hat{\gamma}} 
&=&\frac{(N+1)}{2N}\frac{g^2}{16\pi^2\epsilon}
\left[B^{fgh}_{\alpha\beta\hat{\gamma}}
-2B^{fgh}_{\beta\alpha\hat{\gamma}}\right]\,,
\eeqa
where $\alpha,\beta,\gamma\in\{1,2\}$ (right-handed), 
while $\hat\gamma\in\{\hat{1}=3,\hat{2}=4\}$ (left-handed).
Note that the same results hold with hatted and unhatted indices exchanged.

These relations can be easily diagonalized and the combinations 
which do not mix are given by  
\beqa
(\zeta^{(1)} )^{fgh}_{\{\alpha\alpha\beta\}} &=& 
(\zeta^{(1)} )^{fgh}_{\{\hat\alpha\hat\alpha\hat\beta\}} 
=12\frac{d}{\epsilon}\,, \\
(\zeta^{(1)} )^{f\not=gh}_{[\alpha\beta]\alpha} &=&
(\zeta^{(1)} )^{f\not=gh}_{[\hat\alpha\hat\beta]\hat\alpha} 
= -12\frac{d}{\epsilon}\,,\\
(\zeta^{(1)} )^{fgh}_{\{\alpha\beta\}\hat\gamma} &=&  
(\zeta^{(1)} )^{fgh}_{\{\hat\alpha\hat\beta\}\gamma}
=4\frac{d}{\epsilon}\,, \\
(\zeta^{(1)} )^{f\not=gh}_{[\alpha\beta]\hat\gamma} &=&
(\zeta^{(1)} )^{f\not=gh}_{[\hat\alpha\hat\beta]\gamma} 
= -12\frac{d}{\epsilon}\,,
\eeqa
where $d$ is given by
\beqa
d &\equiv&\frac{1}{32N\pi^2}=\frac{1}{96\pi^2}\,.
\label{defd}
\eeqa
The square bracket denotes antisymmetrization
$[\alpha\beta] =\alpha\beta - \beta\alpha$, 
and curly bracket means
$\{\alpha\beta\} =\alpha\beta + \beta\alpha$, 
$\{\alpha\alpha\beta\} =
\alpha\alpha\beta + \alpha\beta\alpha +\beta\alpha\alpha$.
The totally symmetric case corresponds to the decuplet representation
(for $\Nf=3$) and contains the $\Nf=2\,,I=3/2$ representation.
The antisymmetric case corresponds to the octet representation
(for $\Nf=3$) and contains the $\Nf=2\,,I=1/2$ representation. 

The anomalous dimension at 1-loop is easily obtained from
\beqa
\gamma &=& g^2\gamma^{(1)}+\rmO(g^4)=
\beta_D(g,\epsilon)\frac{\partial\ln\zeta}{\partial g}=
 -2\zeta^{(1)} g^2\epsilon +\rmO(g^4)\,.
\eeqa
Therefore we have
\beqa
\left(\gamma^{(1)}\right)^{fgh}_{\{\alpha\alpha\beta\}}  
= \left(\gamma^{(1)}\right)^{fgh}_{\{\hat\alpha\hat\alpha\hat\beta\}} 
&=& -24d\,,\\
\left(\gamma^{(1)}\right)^{f\not=gh}_{[\alpha\beta]\alpha}  
=\left(\gamma^{(1)}\right)^{f\not=gh}_{[\hat\alpha\hat\beta]\hat\alpha} 
&=& 24d\,,\\
\left(\gamma^{(1)}\right)^{fgh}_{\{\alpha\beta\}\hat\gamma}  
= \left(\gamma^{(1)}\right)^{fgh}_{\{\hat\alpha\hat\beta\}\gamma} &=& 
-8 d\,,\\
\left(\gamma^{(1)}\right)^{f\not=gh}_{[\alpha\beta]\hat\gamma}  
=\left(\gamma^{(1)}\right)^{f\not=gh}_{[\hat\alpha\hat\beta]\gamma} &=& 
24d\,.
\eeqa

\subsection{Anomalous dimensions of 6--quark operators at 1-loop}

In this subsection we consider the renormalization of arbitrary
local gauge invariant 6--quark operator of (lowest) dimension 9.
Any such operator can be written as a linear combination of operators 
\begin{equation}
O_C(x)=B^{F_1,F_2}_{\Gamma_1,\Gamma_2}(x)\equiv 
B^{F_1}_{\Gamma_1}(x) B^{F_2}_{\Gamma_2} (x)=O_A(x)O_B(x)\,, 
\label{sixqops}
\end{equation}
with $A=(\Gamma_1,F_1)$ and $B=(\Gamma_2,F_2)$.  
Note $O_A(x)$ and/or $O_B(x)$ may not be operators with proton or
nucleon quantum numbers and separately may not be diagonally renormalizable 
at one loop. The reason for considering the renormalization 
in more generality is that in principle there may be operators in this class
which occur in the OPE of two nucleon operators at higher order in PT,
but are relevant in the analysis because of their potentially large
anomalous dimensions.

\subsubsection{Linear relations between 6--quark operators}

According to the considerations in subsect.~4.2 the  operators in 
eq.~(\ref{sixqops}) mix only with operators $O_{C'}=O_{A'}O_{B'}$
which preserve the set of flavors and Dirac indices in the chiral basis i.e.
$$
F_1\cup F_2 = F'_1\cup F'_2\,,\,\,\,  
\Gamma_1\cup\Gamma_2 = \Gamma'_1\cup\Gamma'_2\,.
$$
Note however that such operators are not all linearly independent. 
Relations between them follow from a general identity satisfied by
the totally antisymmetric epsilon symbol which for $N$ labels reads
\begin{equation}
N\varepsilon^{a_1\dots a_N}\varepsilon^{b_1\dots b_N}
=\sum_{j,k}\varepsilon^{a_1\dots a_{j-1}b_ka_{j+1}\dots a_N}
\varepsilon^{b_1\dots b_{k-1}a_j b_{k+1}\dots b_N}\,.
\end{equation}
For our special case, $N=3$, this identity implies the following
identities among the 6--quark operators
\beqa
3 B^{F_1,F_2}_{\Gamma_1,\Gamma_2} +
\sum_{i,j=1}^3 B^{(F_1F_2)[i,j]}_{(\Gamma_1,\Gamma_2)[i,j]} = 0\,,
\label{eq:constraint}
\eeqa
where $i$-th index of $abc$ and $j$-th index of $def$ are interchanged 
in $(abc,def)[i,j]$. For example, 
$(\Gamma_1,\Gamma_2)[1,1]=\alpha_2\beta_1\gamma_1,\alpha_1\beta_2\gamma_2$ or
$(\Gamma_1,\Gamma_2)[2,1]= \alpha_1\alpha_2\gamma_1,\beta_1\beta_2\gamma_2$.
Note that the interchange of indices occurs simultaneously for both 
$\Gamma_1,\Gamma_2$ and $F_1,F_2$ in the above formula.
The plus sign in (\ref{eq:constraint}) appears because the quark fields
are Grassmann.

An immediate consequence of the identity is that 
the divergent part of the $\lambda$--dependent contributions, 
calculated from ${\bf T}_1$ in (\ref{eq:1-loopT}), 
must vanish, after the summation over the 9 different contributions 
from quark pairs on the different baryonic parts $A,B$ is taken. 
The $\lambda$--dependent part of the contribution of quark contractions
on the same baryonic parts is compensated by the quark field renormalization. 
Thus the renormalization of the bare 6--quark operator is 
$\lambda$--independent as expected from gauge invariance. 

As an example of identities, we consider the case that 
$\Gamma_1,\Gamma_2=\alpha\alpha\beta,\alpha\beta\beta$ 
($\alpha\not=\beta$ and $F_1,F_2=ffg,ffg$ ($f\not= g$),
the constraint gives
\beqa
&& 3 B_{\alpha\alpha\beta,\alpha\beta\beta}^{ffg,ffg} 
+ (3-2)B_{\alpha\alpha\beta,\alpha\beta\beta}^{ffg,ffg} 
+B_{\alpha\alpha\alpha,\beta\beta\beta}^{fff,fgg}
+(2-1) B_{\alpha\beta\beta,\alpha\alpha\beta}^{fgg,fff} \nn \\
&=&  4 B_{\alpha\alpha\beta,\alpha\beta\beta}^{ffg,ffg}+ 
B_{\alpha\alpha\alpha,\beta\beta\beta}^{fff,fgg}+ 
B_{\alpha\beta\beta,\alpha\alpha\beta}^{fgg,fff}  = 0\,,
\eeqa
where minus signs in the first line come from the property that 
$B_{\Gamma_2,\Gamma_1}^{F_2,F_1}= - B _{\Gamma_1,\Gamma_2}^{F_1,F_2}$. 
There are no further relations among 6--quark operators beyond 
(\ref{eq:constraint}).

\subsubsection{Divergent parts at 1-loop}

We thus need only to calculate the contributions from ${\bf T}_0$, 
which can be classified into the following 4 different 
combinations for a pair of two indices:
\beqa
\left(^f_\alpha\right) \left(^f_\alpha\right) 
&\Rightarrow& -(N+1) \left(^f_\alpha\right) \left(^f_\alpha\right)\,,\\
\left(^f_\alpha\right) \left(^f_\beta\right) 
&\Rightarrow& (1-2N) \left(^f_\alpha\right) \left(^f_\beta\right)  
+(N-2)\left(^f_\beta\right) \left(^f_\alpha\right)\,,\\
\left(^{f_1}_\alpha\right) \left(^{f_2}_\alpha\right) 
&\Rightarrow& -\left(^{f_1}_\alpha\right) \left(^{f_2}_\alpha\right) 
-N\left(^{f_2}_\alpha\right) \left(^{f_1}_\alpha\right)\,,\\
 \left(^{f_1}_\alpha\right) \left(^{f_2}_\beta\right) 
&\Rightarrow& \left\{\left(^{f_1}_\alpha\right) \left(^{f_2}_\beta\right)  -2\left(^{f_1}_\beta\right)
\left(^{f_2}_\alpha\right)  \right\} +N \left\{\left(^{f_2}_\beta\right) \left(^{f_1}_\alpha\right)  -2\left(^{f_2}_\alpha\right)
\left(^{f_1}_\beta\right)  \right\}\,,
\eeqa
where $f\not= g$ and  $\alpha\not= \beta \in (1,2)$ (Right) or 
$\in (3,4)$ (Left).

The computation can be made according to the following steps:

i.) Select the total flavor content e.g. $3f+3g$ or $4f+2g$ ($f\ne g$).
These are the only cases we will consider since in this paper 
we are mainly restricting attention to baryon operators with $\Nf=2$, 
but the approach is also applicable to more general cases ($\Nf>2$).

ii.) Given a flavor content classify all the possible sets
of Dirac labels in the chiral basis e.g.  $111223,112234,...$
It is obvious from the rules above 
that some have equivalent renormalization at 1-loop e.g. 
$111223$ and $112223$ with $1\leftrightarrow2$,
and also those with hatted and unhatted indices exchanged e.g. $111223$ 
and $133344$.

iii.) For given flavor and Dirac sets generate all possible operators
\footnote{recall the single baryon operators are symmetric under 
exchange of pairs of indices}. Then
generate all gauge identities between them and determine a maximally 
independent (basis) set $\{\mathcal{S}_i\}$. 

iv.) Compute the divergent parts of the members of the independent 
basis:
\be
\Gamma_i^{\mathrm{div}}=\frac{1}{2\epsilon}\gamma_{ij}\mathcal{S}_j\,.
\end{equation}

v.) Finally compute the eigenvalues and corresponding eigenvectors of 
$\gamma^T$ to determine the operators which renormalize diagonally at 1-loop.

An example of the procedure is given in Appendix~A. Some of the
steps are quite tedious if carried out by hand. e.g. in the case $3f+3g$
and Dirac indices $112234$ there are initially 68 operators in step iii.
with 38 independent gauge identities, and hence an independent basis of 30
operators. However all the steps above can be easily implemented
in an algebraic computer program using MATHEMATICA or MAPLE.

If the quarks $f,g$ belong to an iso-doublet e.g. we identify $f$ with $u$
having $I_3=1/2$ and $g$ with $d$ ($I_3=-1/3$), then if an eigenvalue
is non-degenerate the corresponding eigenvector 
belongs to a certain representation of the isospin group. 
If the eigenvalue is degenerate then linear combinations of them
belong to definite representations.
For the $3f+3g$ case they can have $I=0,1,2,3$. 
Eigenvectors with $I=0,2$ are odd under the interchange $f\leftrightarrow g$ 
and those with $I=1,3$ are even. The operators in the case $4f+2g$ have
$I_3=1$ and hence have $I=1,2,3$. The eigenvectors in this case 
can be obtained from those of the $3f+3g$ case by applying the isospin 
raising operator. 

The complete list of eigenvalues and possible isospins are given in 
Tables~\ref{T3f3ga}-\ref{T3f3gc} in Appendix~A. 
Here we summarize the most important results.

1) For the $3f+3g$ (and $4f+2g$) cases all eigenvalues 
$\gamma_j\le 48d=2\gamma_N$, where $\gamma_N$ is the 1-loop anomalous dimension of the nucleon (3--quark) operator. 
We have not found an elegant way of proving this other than computing
all cases explicitly.

2) It is easy to construct eigenvectors with eigenvalue $2\gamma_N$ 
e.g. operators of the form 
$B^{ffg}_{\alpha[\beta,\alpha]}
B^{ggf}_{\hat{\alpha}[\hat{\beta},\hat{\alpha]}}$ since
there is no contribution from diagrams where the gluon line joins quarks
in the different baryonic parts.

3) Operators with higher isospin generally have smaller eigenvalues.

\subsubsection{Decomposition of two-nucleon operators}
Since 
\beqa
C \gamma_5 &=& \left(\begin{array}{cccc}
0 & -1 & 0 & 0 \\
1 & 0 & 0 & 0 \\
0 & 0 & 0 & -1 \\
0 & 0 & 1 & 0 \\
\end{array}
\right)\,, 
\eeqa
in the chiral representation, the nucleon operator is written as
\beqa
B_{\alpha}^f &=& B_{\alpha+\hat\alpha, [2,1]}^{ffg}
+B_{\alpha+\hat\alpha, [\hat 2,\hat 1]}^{ffg}
\eeqa
where $\alpha = 1,2$, $\hat \alpha =\alpha+2$, and $f\not= g$.
This has anomalous dimension $\gamma_N=24d$.

We then consider two independent 6--quark operators occurring in the OPE
at tree level which decomposed as follows.
The  spin-singlet ($S=0$) and isospin-triplet ($I=1$) operator is decomposed as
\beqa
B^{ffg}_{\alpha+ \hat\alpha, [\beta,\alpha]+ [\hat \beta,\hat \alpha]}
B^{ffg}_{\beta+ \hat\beta, [\beta,\alpha]+ [\hat \beta,\hat \alpha]}
&=& B_I^{01} + B_{II}^{01} + B_{III}^{01} + B_{IV}^{01} + B_V^{01} + B_{VI}^{01}
\eeqa
where
\beqa
B_I^{01} &=&B^{ffg}_{\alpha [\beta,\alpha]}B^{ffg}_{\beta [\beta,\alpha]} +
B^{ffg}_{\hat\alpha [\hat\beta,\hat\alpha]}B^{ffg}_{\hat\beta[\hat\beta,\hat\alpha]}\,,\\
B_{II}^{01} &=&
B^{ffg}_{\alpha [\beta,\alpha]}B^{ffg}_{\beta [\hat\beta,\hat\alpha]}
+B^{ffg}_{\alpha [\hat\beta,\hat\alpha]}B^{ffg}_{\beta [\beta,\alpha]}  
+B^{ffg}_{\hat\alpha [\hat\beta,\hat\alpha]}B^{ffg}_{\hat\beta [\beta,\alpha]} 
+B^{ffg}_{\hat\alpha [\beta,\alpha]}B^{ffg}_{\hat\beta[\hat\beta,\hat\alpha]}\,, 
\\
B_{III}^{01} &=& B^{ffg}_{\alpha [\beta,\alpha]}B^{ffg}_{\hat\beta [\beta,\alpha]} 
+B^{ffg}_{\hat \alpha [\beta,\alpha]}B^{ffg}_{\beta [\beta,\alpha]}
+B^{ffg}_{\hat\alpha [\hat\beta,\hat\alpha]}B^{ffg}_{\beta [\hat\beta,\hat\alpha]}
+ B^{ffg}_{\alpha [\hat\beta,\hat\alpha]}B^{ffg}_{\hat\beta[\hat\beta,\hat\alpha]}\,,
\\
B_{IV}^{01}
&=&B^{ffg}_{\alpha [\hat\beta,\hat\alpha]}B^{ffg}_{\beta [\hat\beta,\hat\alpha]}
+B^{ffg}_{\hat\alpha [\beta,\alpha]}B^{ffg}_{\hat\beta[\beta,\alpha]}\,,\\
B_V^{01}
&=&B^{ffg}_{\alpha [\hat\beta,\hat\alpha]}B^{ffg}_{\hat\beta [\beta,\alpha]}
+B^{ffg}_{\hat\alpha [\beta,\alpha]}B^{ffg}_{\beta[\hat\beta,\hat\alpha]}\,,\\
B_{VI}^{01}
&=&B^{ffg}_{\alpha [\beta,\alpha]}B^{ffg}_{\hat\beta [\hat\beta,\hat\alpha]}
+B^{ffg}_{\hat\alpha [\hat\beta,\hat\alpha]}B^{ffg}_{\beta[\beta,\alpha]}\,,
\eeqa
where $\alpha\not= \beta$.
In the above we do not have to calculate all contributions. Some of them are obtained from
interchanges under $(1,2)\leftrightarrow (3,4)$ or $(1,3)\leftrightarrow (2,4)$.

Similarly the spin-triplet ($S=1$) and isospin-singlet ($I=0$) 
operator is decomposed as 
\beqa
B^{ffg}_{\alpha+ \hat\alpha, [\beta,\alpha]+ [\hat \beta,\hat \alpha]}
B^{ggf}_{\alpha+ \hat\alpha, [\beta,\alpha]+ [\hat \beta,\hat \alpha]}
&=& B_I^{10} + B_{II}^{10} + B_{III}^{10} + B_{IV}^{10} 
+ B_V^{10} + B_{VI}^{10}\,,
\eeqa
where
\beqa
B_I^{10}&=&
B^{ffg}_{\alpha [\beta,\alpha]}B^{ggf}_{\alpha [\beta,\alpha]} +
B^{ffg}_{\hat\alpha [\hat\beta,\hat\alpha]}B^{ggf}_{\hat\alpha[\hat\beta,\hat\alpha]}\,,
\\
B_{II}^{10}&=&
B^{ffg}_{\alpha [\beta,\alpha]}B^{ggf}_{\alpha [\hat\beta,\hat\alpha]}
+B^{ffg}_{\alpha [\hat\beta,\hat\alpha]}B^{ggf}_{\alpha [\beta,\alpha]}  
+B^{ffg}_{\hat\alpha [\hat\beta,\hat\alpha]}B^{ggf}_{\hat\alpha[\beta,\alpha]} 
+B^{ffg}_{\hat\alpha [\beta,\alpha]}B^{ggf}_{\hat\alpha[\hat\beta,\hat\alpha]}\,, 
\\
B_{III}^{10}&=& 
B^{ffg}_{\alpha [\beta,\alpha]}B^{ggf}_{\hat\alpha [\beta,\alpha]} 
+B^{ffg}_{\hat \alpha [\beta,\alpha]}B^{ggf}_{\alpha [\beta,\alpha]}
+B^{ffg}_{\hat\alpha [\hat\beta,\hat\alpha]}B^{ggf}_{\alpha [\hat\beta,\hat\alpha]}
+B^{ffg}_{\alpha [\hat\beta,\hat\alpha]}B^{ggf}_{\hat\alpha[\hat\beta,\hat\alpha]}\,,
\\
B_{IV}^{10}&=&
B^{ffg}_{\alpha [\hat\beta,\hat\alpha]}B^{ggf}_{\alpha [\hat\beta,\hat\alpha]}
+B^{ffg}_{\hat\alpha [\beta,\alpha]}B^{ggf}_{\hat\alpha [\beta,\alpha]}\,, 
\\
B_V^{10}&=&
B^{ffg}_{\alpha [\hat\beta,\hat\alpha]}B^{ggf}_{\hat\alpha [\beta,\alpha]}
+B^{ffg}_{\hat\alpha [\beta,\alpha]}B^{ggf}_{\alpha[\hat\beta,\hat\alpha]}\,, 
\\
B_{VI}^{10}&=&
B^{ffg}_{\alpha [\beta,\alpha]}B^{ggf}_{\hat\alpha [\hat\beta,\hat\alpha]}
+B^{ffg}_{\hat\alpha [\hat\beta,\hat\alpha]}B^{ggf}_{\alpha[\beta,\alpha]}\,.
\eeqa
Again a half of the above 1-loop contributions can be obtained from others 
by the interchange $(1,2)\leftrightarrow (3,4)$ or $f\rightarrow g$.
 
\subsubsection{Results for anomalous dimensions}

It is very important to note here that operators $B_{VI}^{SI}$ for 
both cases ($SI =01$ and $10$) have the maximal anomalous dimension at 1-loop, 
since as noted in point 2) above, no 1-loop correction from ${\bf T}_0$ 
joining quarks from the two baryonic components
exists for $B_{\alpha\beta\gamma,\hat\alpha'\hat\beta'\hat\gamma'}^{F_1,F_2}$ 
type of operators.
Therefore we always have some operators with $\beta_{A} = 0$
which dominate in the OPE at short distance.

The 1-loop corrections $\Gamma^{(1)}$ to 6--quark operators $B^{SI}$ 
are computed in appendix~\ref{appendixA} and are summarized as:
\beqa
\left(\Gamma_{I}^{01}\right)^{(1)} &=& -12\frac{d}{\epsilon}B_I^{01}\,,\quad
\left(\Gamma_{II}^{01}\right)^{(1)} = 12\frac{d}{\epsilon}B_{II}^{01}\,,\quad
\left(\Gamma_{III}^{01}\right)^{(1)} = 0\,,\nn \\
\left(\Gamma_{IV}^{01}\right)^{(1)} &=& 0\,, \
\left(\Gamma_{V}^{01}\right)^{(1)} = 6\frac{d}{\epsilon} B_{V}^{01}  
+ 6\frac{d}{\epsilon} B_{VI}^{01}\,,\
\left(\Gamma_{VI}^{01}\right)^{(1)} = 24\frac{d}{\epsilon}B_{VI}^{01}\,, 
\eeqa
for $SI = 01$.  
The last two results can be written as
\beqa
\left(\Gamma_{V'}^{01}\right)^{(1)} &=& 6\frac{d}{\epsilon} B_{V'}^{01}\,,
\quad
\left(\Gamma_{VI'}^{01}\right)^{(1)} = 24\frac{d}{\epsilon}B_{VI'}^{01}\,,\,, 
\eeqa
where
\beqa
B_{V'}^{01} &=& B_V^{01} - \frac{1}{3} B_{VI}^{01}\,, \qquad
B_{VI'}^{01} = B_{VI}^{01}\,.
\eeqa
Similarly we have for $SI=10$
\beqa
\left(\Gamma_{I}^{10}\right)^{(1)} &=& -4\frac{d}{\epsilon}B_I^{10}\,,\quad
\left(\Gamma_{II}^{10}\right)^{(1)} = 20\frac{d}{\epsilon}B_{II}^{10}\,,\quad
\left(\Gamma_{III}^{10}\right)^{(1)} = 0\,,\nn \\
\left(\Gamma_{IV}^{10}\right)^{(1)} &=& 8\frac{d}{\epsilon} B_{IV}^{10}\,,\quad
\left(\Gamma_{V'}^{10}\right)^{(1)} = 6\frac{d}{\epsilon}B_{V'}^{10}\,,\quad
\left(\Gamma_{VI'}^{10}\right)^{(1)} = 24\frac{d}{\epsilon}B_{VI'}^{10}\,,
\eeqa
where
\beqa
B_{V'}^{10} &=& B_V^{10} - \frac{1}{3} B_{VI}^{10}\,, \qquad
B_{VI'}^{10} = B_{VI}^{10}\,.
\eeqa

Denoting the eigenvalues of the anomalous dimension matrix by $\gamma_C$, 
we give the values of $\gamma^{SI}$ defined by
\begin{equation}
\gamma_C-2\gamma_N= 2d \gamma^{SI}\,,
\label{gammaSI}
\end{equation}
in table~\ref{tab:gamma} ($N=3$),
which shows, in both cases, that the largest value is zero 
while others are all negative.
The case 2 in sect.~\ref{sec:3dim} is realized: $\beta_C = 0$ and
\beqa
\beta_{C^\prime} =\beta^{01}_0
&=&  -\frac{6}{33-2\Nf} \qquad \mbox{for} \ S=0\,, I=1\,, \label{beta01}\\
\beta_{C^\prime} =\beta^{10}_0
&=&  -\frac{2}{33-2\Nf} \qquad \mbox{for} \ S=1\,, I=0\,.\label{beta10}
\eeqa

\begin{table}[tb]
\caption{The value of $\gamma^{SI}$ (defined in (4.72)) for 
each eigen operator in the $SI=01$ and $SI=10$ states. } 
\label{tab:gamma}
\begin{center}
\begin{tabular}{|c|cccccc|}
\hline
 & $I$ & $II$ & $ III$ &$ IV$ & $V'$ & $VI'$ \\
\hline
$\gamma^{01}$ & $-36$ & $-12$ & $-24$ & $-24$ & $-18$ & $0$ \\
$\gamma^{10}$ & $-28$ & $-4$ & $-24$ & $-16$ & $-18$ & $0$ \\
\hline
\end{tabular}
\end{center}
\end{table}




\section{Short distance behavior of the nucleon potential}
We consider the following structure of the potential.
\beqa
V(\vec y) &=& V_0(r) + V_\sigma(r) ({\vec \sigma_1\cdot\vec \sigma_2}) 
+ V_T(r) S_{12} 
+\rmO(\nabla) 
\eeqa  
where $r=\vert \vec y\vert$, and 
\beqa
S_{12} &=& 3 ({\vec\sigma_1\cdot \hat{\vec y}})
({\vec\sigma_2\cdot \hat{\vec y}}) 
- ({\vec \sigma_1\cdot\vec\sigma_2}),
\quad \hat {\vec y} = \frac{\vec y}{\vert \vec y \vert}
\label{S_12}
\eeqa
is the tensor operator. Here $\vec\sigma_i$ acts on the spin labels of
the $i^{\rm th}$ nucleon.

Since, as shown in the previous section,
6--quark operators appeared at tree level in the OPE of 
NN which have the largest and the second largest
anomalous dimensions,
we mainly consider 6--quark operators at tree level in the OPE,
which is written as
\beqa
B_\alpha^f (x+y/2) B_\beta^g(x-y/2) &\simeq& c_{VI} B_{VI,\alpha\beta}^{fg}(x)
+ c_{II} (-\log r)^{\beta^{SI}_0}  B_{II,\alpha\beta}^{fg}(x) + \cdots
\label{eq:ope_full}
\eeqa
where $c_{VI}$ and $c_{II}$ are some constants, and $\cdots$ represents other contributions, which are less singular than the first two at short distance.
(We here write the spinor and flavor indices $\alpha, \beta$ and $f,g$ 
explicitly for later use.)
The anomalous dimensions $\beta_0^{SI}$ are given in (\ref{beta01}) and
(\ref{beta10}).

\subsection{Potential for $S=0$ and $I=1$ states}
In the case that $S=0$ and $I=1$, we take $\alpha\not=\beta$ and $f=g$ in eq.(\ref{eq:ope_full}), whose leading contributions couple only to the $J=L=0$ state, which is given by
\beqa
\vert E \rangle = \vert L_z=0, S_z=0,I_z=1\rangle_{L=0,S=0,I=1} 
= \vert 0, 0,1\rangle_{0,0,1}\,.
\eeqa 
The relevant matrix elements are given by 
\beqa
c_{VI}\langle 0 \vert B_{VI,\alpha\beta}^{fg}\vert 0, 0,1\rangle_{0,0,0} &=& A_{VI}^0 Y_0^0 [\alpha\beta]\{fg\}_1,\\
c_{II}\langle 0 \vert B_{II,\alpha\beta}^{fg}\vert 0, 0,1\rangle_{0,0,0} &=& A_{II}^0 Y_0^0 [\alpha\beta]\{fg\}_1,
\eeqa
where $A_{II}^0$ and $A_{VI}^0$ are non-perturbative constants, $Y_L^{L_z}$ is a spherical harmonic function, $[\alpha\beta] =(\delta_{\alpha 1}\delta_{\beta 2}-\delta_{\beta 1}\delta_{\alpha 2})/\sqrt{2}$ represents the $(S,S_z)=(0,0)$ component, and $\{fg\}_1 = \delta_{f1}\delta_{g1}$ 
corresponds to isospin $(I,I_z) =(1,1)$.
With the notation that $\phi(^1S_0,J_z=0)^{II_z=11} = Y_0^0 [\alpha\beta]\{fg\}_1$, 
the wave function at short distance is dominated by
\beqa
\varphi_E^{^1S_0} (y) &=&\langle 0 \vert B_\alpha^f (x+y/2) B_\beta^g(x-y/2) \vert 0,0,1\rangle_{0,0,1}\nn \\
&\simeq& \left( A_{VI}^0 + A_{II}^0(-\log r)^{\beta^{01}_0}\right) \phi(^1S_0, 0)^{11} 
+\cdots
\eeqa
from which we obtain
\beqa
\frac{\nabla^2}{2m} \varphi_E^{ ^1S_0} (y) 
&\simeq& \frac{(-\log r)^{\beta^{01}_0-1}}{r^2}
\frac{-\beta^{01}_0 A_{II}^0 }{m_N} \phi(^1S_0, 0)^{11} +\cdots
\eeqa
where $m=m_N/2$ is the reduced mass of the two nucleon system. 

Since $S_{12}$ is zero on $\phi(^1S_0, 0)^{11} $,
we have
\beqa
V_c^{01}(r) \varphi_E^{^1S_0} (y) &\simeq & 
\frac{(-\log r)^{\beta^{01}_0-1}}{r^2}
\frac{-\beta^{01}_0 A_{II}^0 }{m_N} \phi(^1S_0, 0)^{11} +\cdots
\eeqa
where $V_c^{01}(r) = V_0(r) -3 V_\sigma (r)$. We therefore obtain
\beqa
V_c^{01}(r)&\simeq& F^{01}(r)\frac{A_{II}^0}{ A_{VI}^0}\,.
\eeqa
where
\beqa
F^{SI}(r) &=& \frac{-\beta^{SI}_0 (-\log r)^{\beta^{SI}_0-1}}{m_N r^2}\,.
\eeqa
The potential diverges as $F^{01}(r)$ in the $r\rightarrow 0$ limit, which is a little weaker than $r^{-2}$.

\subsection{Potential for $S=1$ and $I=0$}
We here consider the spin-triplet and isospin-singlet state($S=1$ and $I=0$).
Since $I=0$ and $I_z=0$ in this case, we drop indices $I$ and $I_z$ unless necessary. 
In this case,  the leading contributions  in eq.(\ref{eq:ope_full}) 
couple only to the $J=1$ state,
which is given by 
\beqa
\vert E \rangle &=& \vert ^3S_1,J_z=1\rangle + x  \vert^3D_1,J_z=1\rangle\,,
\eeqa
where 
\beqa
\vert ^3 S_1, J_z=1\rangle &=& \vert L_z=0, S_z=1\rangle_{L=0,S=1}\,,\\
\vert ^3 D_1, J_z=1\rangle &=& \frac{1}{\sqrt{10}}\left[ \vert 0, 1\rangle
-\sqrt{3} \vert 1,0\rangle
+ \sqrt{6}\vert 2, -1\rangle\right]_{L=2,S=1}\,,
\eeqa
$x$ is the mixing coefficient, which is determined by QCD dynamics, and
$^{2S+1} L_J$ specifies quantum numbers of the state.

Relevant matrix elements are given by
\beqa
c_i
\left\langle 0 \left\vert  B_{i,\alpha\beta}^{fg}  \right\vert {}^3S_1, 1\right\rangle
&=& B_i^0   \phi\left({}^3S_1\right)\,, \qquad
 c_i 
\left\langle 0\left\vert  B_{i,\alpha\beta}^{fg}  \right\vert {}^3D_1, 1\right\rangle
= 0\,,  
\eeqa
for $i=II$ and $VI$, 
where $B_i^0$ are non-perturbative constants, and 
\beqa
\phi\left({}^3S_{1}\right) &=& Y_0^0(\theta,\phi)\{\alpha\beta\}_1[fg]\,,
\eeqa
with $\{\alpha\beta\}_1=\delta_{\alpha 1}\delta_{\beta 1}$.

Using the above results, we have
\beqa
\varphi_E^{J=1}(y) &\simeq&   \left\{ B_{VI}^0 +   (-\log r)^{\beta^{10}_{0}} B_{II}^0\right\}
\phi\left({}^3S_{1}\right) ,
\eeqa
By applying $\nabla^2$ we obtain
\beqa
\frac{\nabla^2}{2m}\varphi_E^{J=1} (y) &\simeq& \frac{-\beta^{10}_{}}{m_N} \frac{(-\log r)^{\beta^{10}_{0}-1}}{ r^2} 
B_{II}^0\phi\left({}^3S_{1}\right) \,, 
\label{eq:derV}
\eeqa 
while
\beqa
V(y) \varphi_E^{J=1}(y) &\simeq&  B_{VI}^0 V_c^{10}(r)  \phi\left({}^3S_{1}\right)
+  2\sqrt{2} V_T(r) B_{VI}^0 \phi\left({}^3D_{1}\right)\,,
\label{eq:genV}
\eeqa
where we use the  formula in the appendix~\ref{appendixB}, 
$V_c^{10}(r) = V_0^0(r) + V_\sigma^0 (r)$, and
\beqa
\phi\left({}^3D_{1}\right) &=& \frac{1}{\sqrt{10}}\left[ Y_{2}^0\{\alpha\beta\}_1-\sqrt{3} Y_{2}^{1}\{\alpha\beta\}_0 +\sqrt{6}Y_{2}^{2}\{\alpha\beta\}_{-1}\right] [fg]\,,
\eeqa
with $\{\alpha\beta\}_{-1}=\delta_{\alpha 2}\delta_{\beta 2}$ and 
$\{\alpha\beta\}_0=(\delta_{\alpha 1}\delta_{\beta 2}+\delta_{\beta1}\delta_{\alpha 2})/\sqrt{2}$.

By comparing eq.(\ref{eq:derV}) with eq.(\ref{eq:genV}), we have
\beqa
V_c^{10}(r) &\simeq& F^{10}(r) \frac{B_{II}^0}{B_{VI}^0}\,, \qquad
V_T(r) \simeq 0\,.
\eeqa
This shows that the central potential $V_c^{10}(r)$ diverges as $F^{10}(r)$ 
in the $r\rightarrow 0$ limit, which is a little weaker than $1/r^2$, 
while the tensor potential $V_T(r)$ becomes zero in this limit at the tree level in the OPE.

\subsection{Higher order in the OPE and the tensor potential}
While the short distance behavior  of the central potential is determined by the OPE at the tree level,  the determination of the tensor potential at short distance requires 
the OPE at higher order, whose relevant contribution is given by
\beqa
B_\alpha^f (x+y/2) B_\beta^g(x-y/2) &\simeq& c_{VI} B_{VI,\alpha\beta}^{fg}(x)
+ c_{II} (-\log r)^{\beta^{SI}_0}  B_{II,\alpha\beta}^{fg}(x) \nn \\
&+& c_T (-\log r)^{\beta^{SI}_T} \frac{[ y^k y^l]}{r^2}B^{fg,kl}_{T,\alpha\beta}(x) +\cdots
\cdots
\label{eq:ope_high}
\eeqa
where $[y^ky^l] = y^k y^l - r^2\delta^{kl}/3$, and we assume that the third term with
the tensor-type operator $B_{T,\alpha\beta}^{fg,kl}$ appears first 
at $\ell_T^{SI} ( > 0) $ loop of the perturbative expansion in the OPE. 
Therefore, with this assumption, we have
\beqa
\beta_T^{SI} &=& - \ell_T^{SI} + \frac{\Delta_T^{SI} - 24}{2(33-2N_f)}
\eeqa
where $\Delta^{SI}_T =\gamma_T/(2d)$ 
with $\gamma_T$ being the anomalous dimension of the operator 
$B_{T,\alpha\beta}^{fg,kl}$. The calculation of anomalous dimensions 
for all 6--quark operators in the previous section shows that 
$\Delta_T \le 24$, so that $ \beta_T^{SI} < \beta^{SI}_0 < 0$.

An extra matrix element we need is given as
\beqa
x\, c_T \frac{[ y^ky^l]}{r^2}
\left\langle 0\left\vert  B_{T,\alpha\beta}^{fg,kl}  \right\vert {}^3D_{1}, 1\right\rangle
&=& B_T  \phi\left({}^3D_{1}\right)\,, 
\eeqa
where $B_T$ is a further non-perturbative constant. 

Using the above results, we have for $(S,I) = (1,0)$ 
\beqa
\varphi_E^{J=1}(y) &\simeq& \left\{ B_{VI}^0 +   (-\log r)^{\beta^{10}_{0}} B_{II}^0\right\}
\phi\left({}^3S_{1}\right) 
+ 
(-\log r)^{\beta^{10}_T} B_T 
\phi\left({}^3D_{1}\right)\,.
\eeqa
By applying $\nabla^2$ we obtain
\beqa
\frac{\nabla^2}{2m}\varphi_E^{J=1} (y) &\simeq& \frac{-\beta^{10}_{0}}{m_N} \frac{(-\log r)^{\beta^{10}_{0}-1}}{ r^2} 
B_{II}^0\phi\left({}^3S_{1}\right)  \nn \\
&+& \frac{-6}{m_N} \frac{(-\log r)^{\beta^{10}_{T}}}{ r^2} 
B_{T} \phi\left({}^3D_{1}\right)\,. 
\eeqa 

From $V(y)\varphi_E^{J=1}(y) = (E + \nabla^2/(2m) )\varphi_E^{J=1}(y) $, we obtain
 \beqa
V_c^{10}(r)  
&\simeq&  F^{10}(r)  \frac{B_{II}^0}{B_{VI}^0}\,,  \\
 V_T(r)  
&\simeq&   F_{T}^{10}(r) \frac{-3 B_{T}}{\sqrt{2} B_{VI}^0}\,,  
\label{eq:eq2a}
\eeqa
where
\beqa
F_T^{10}(r) &=& \frac{(-\log r)^{\beta^{10}_T}}{m_N r^2}\,.
\eeqa
This shows that the central potential $V_c(r)$ diverges as $F^{10}(r)$ 
in the $r\rightarrow 0$ limit, which is a little weaker than $1/r^2$, 
while the tensor potential $V_T(r)$ diverges as $F^{10}_T(r)$ in this limit,
which is not stronger than $F^{10}(r)$ since  $\beta^{10}_0 -1 \ge \beta^{10}_T$.

\section{Evaluation of matrix elements}

We rewrite 3--quark operators in terms of left- and right- handed component:
\beqa
B_{X,\alpha}^f &=& B_{\alpha\beta\gamma}^{fgh} 
(C\gamma_5 P_X)_{\beta\gamma}(i\tau_2)_{gh}\,,
\\
B_{XY,\alpha}^f &=& (P_X)_{\alpha\beta} B_{Y,\beta}^f\,, 
\eeqa
for $X,Y = R$ or $L$. In terms of these we have 
\beqa
B_{\alpha}^f B_{\beta}^g 
&=& \left[B_I + B_{II} + B_{III} + B_{IV} + B_{V} 
+ B_{VI}\right]^{fg}_{\alpha\beta}\,,
\eeqa
where
\beqa
(B_I)^{fg}_{\alpha\beta} &=& 
\left[ B_{RR}B_{RR}+B_{LL}B_{LL}\right]^{fg}_{\alpha\beta}\,,\\
(B_{II})^{fg}_{\alpha\beta} &=& 
\left[ B_{RR}B_{RL}+B_{RL}B_{RR}
+B_{LL}B_{LR}+B_{LR}B_{LL}\right]^{fg}_{\alpha\beta}\,,\\
(B_{III})^{fg}_{\alpha\beta} &=& 
\left[ B_{RR}B_{LR} +B_{LR}B_{RR}+B_{LL}B_{RL}+B_{RL}B_{LL}\right]^{fg}_{\alpha\beta}\,,\\
(B_{IV})^{fg}_{\alpha\beta} &=& 
\left[ B_{RL}B_{RL}+B_{LR}B_{LR}\right]^{fg}_{\alpha\beta}\,,\\
(B_{V})^{fg}_{\alpha\beta} &=& 
\left[ B_{RL}B_{LR} +B_{LR}B_{RL}\right]^{fg}_{\alpha\beta}\,,\\
(B_{VI})^{fg}_{\alpha\beta} &=& 
\left[ B_{RR}B_{LL} +B_{LL}B_{RR}\right]^{fg}_{\alpha\beta}\,.
\eeqa
Note that we take $(\vec x , t) =(\vec 0,0)$ in the above operators.
We need to know
\beqa
\langle 0 \vert (B_i)^{fg}_{\alpha\beta} \vert 2N, E\rangle
\eeqa
for $i=II,VI$.

For $f\not= g$, Lorentz covariance leads to 
\beqa
\langle 0 \vert B_X^f B_Y^g  \vert  2{\rm N}, E\rangle
&=& \sum_{A,B=R,L} C_{XY}^{AB}(s)  P_A u (\vec p, \sigma_1) P_B u(-\vec p,\sigma_2),
\label{eq:FF}
\eeqa
where $s= E^2 = 4 \sqrt{\vec p^2 +m_N^2}$ with the total energy $E$ in the center of mass frame, $\sigma_i$ ($i=1,2$) is the spin of the $i$-th nucleon, and $C_{XY}^{AB}$ is an unknown function of $s$.  Using invariance of QCD under the parity transformation $ P B_X P^{-1}  = \gamma_4 B_{\bar X} $ 
where $\bar R = L$ and $\bar L = R$, we rewrite  eq.(\ref{eq:FF}) as
\beqa
(\ref{eq:FF}) &=&
\langle 0 \vert  P B_X^f B_Y^g P^{-1} P  \vert  2{\rm N}, E\rangle
= \sum_{A.B} C_{\bar X \bar Y}^{AB} P_{\bar A} \gamma_4 u(-\vec p,\sigma_1) P_{\bar B}
\gamma_4 u (\vec p, \sigma_2) \nn \\
&=& \sum_{A,B} C_{\bar X \bar Y}^{\bar A\bar B} P_{A} u(\vec p,\sigma_1) P_{B}
u (-\vec p, \sigma_2),
\eeqa
where $\gamma_4 u(-\vec p,\sigma_1) = u(\vec p,\sigma_1)$ is used. The above relation implies $C_{\bar X \bar Y}^{\bar A\bar B} = C_{XY}^{AB}$. Using this property for the unknown functions $C_{XY}^{AB}$, we have
\beqa
\langle 0 \vert (B_{II})^{fg\pm gf}_{\alpha\beta} \vert 2{\rm N}, E\rangle
&=& C_{RL+LR}^{RR,\pm}\{ (P_R\otimes P_R + P_L\otimes P_L) u(\vec p,\sigma_1)
u(-\vec p,\sigma_2) \}_{\alpha\beta\mp\beta\alpha}
\eeqa 
and
\beqa
\langle 0 \vert (B_{VI})^{fg\pm gf}_{\alpha\beta} \vert 2{\rm N}, E\rangle
&=& C_{RL}^{RL,\pm}\{ (P_R\otimes P_L + P_L\otimes P_R) u(\vec p,\sigma_1)  u(-\vec p,\sigma_2) \}_{\alpha\beta\mp\beta\alpha}
\eeqa 
 
Taking $\vec p =(0,0, p_z> 0)$ and Dirac representation 
for $\gamma$ matrices \cite{IZ}, we have
\beqa
u(\pm\vec p, +) &= \dfrac{1}{\sqrt{E_N+ m_N}}\left(
\begin{array}{c}
E_N + m_N \\
0 \\
\mp p_z \\
0 \\
\end{array}
\right)\, 
u(\pm \vec p, -) &= \frac{1}{\sqrt{E_N+ m_N}}\left(
\begin{array}{c}
0 \\
E_N + m_N \\
0 \\
\pm p_z \\
\end{array}
\right)
\eeqa
where $E_N=\sqrt{\vec p^2+m_N^2}$.
For $I=1$ ( $fg+gf$) and $ S=0$ ($ \sigma_1=+$ and $\sigma_2=-$ )
the above explicit form for the spinors gives
\beqa
\{ (P_R\otimes P_R + P_L\otimes P_L) u(\vec p,+)  u(-\vec p,-) \}_{12-21}
&=& E_N\,, \\ 
\{ (P_R\otimes P_L + P_L\otimes P_R) u(\vec p,+)  u(-\vec p,-) \}_{12-21}
&=& m_N\, ,
\eeqa
while, for $I=0$ ( $fg-gf$) and $ S=1$ ($ \sigma_1=+$ and $\sigma_2=+$ ),
we have
\beqa
 \{ (P_R\otimes P_R + P_L\otimes P_L) u(\vec p,+)  u(-\vec p,+) \}_{11}
&=& m_N\,,  \\
\{ (P_R\otimes P_L + P_L\otimes P_R) u(\vec p,+)  u(-\vec p,+) \}_{11}
&=& E_N\, .
\eeqa

We finally obtain
\beqa
\frac{\langle 0 \vert (B_{II})^{fg+gf}_{12} \vert 2{\rm N}, E\rangle }
 {\langle 0 \vert (B_{VI})^{fg+gf}_{12} \vert 2{\rm N}, E\rangle } 
&=& \frac{ E_N}{m_N}\frac{C_{RL+LR}^{RR,+}(s)}{C_{RL}^{RL,+}(s)}
\eeqa
for $fg+gf$ and $(\sigma_1,\sigma_2)=(+,-)$ ( $^1S_0$ ), and
\beqa
\frac{\langle 0 \vert (B_{II})^{fg-fg}_{11} \vert 2{\rm N}, E\rangle }
 {\langle 0 \vert (B_{VI})^{fg-fg}_{11} \vert 2{\rm N}, E\rangle } &=& 
\frac{m_N}{E_N}
 \frac{C_{RL+LR}^{RR,-}(s)}{C_{RL}^{RL,-}(s)}
\eeqa
for $fg-gf$ and $(\sigma_1,\sigma_2)=(+,+)$ ( $^3S_1$ ), where $s=4E_N^2$.

Unfortunately, we can not determine the sign of the ratio for these matrix elements.
As a very crude estimation, we consider the non-relativistic expansion for constituent quarks whose mass $m_Q$ is given by $m_Q=m_N/3$. 
In the large $m_Q$ limit, $\gamma_4 q_0 = q_0$ and $\gamma_4 u_0 = u_0$, where a subscript $0$ for $q$ and $u$ means the 0-th order in the non-relativistic expansion.
In this limit, it is easy to show $C_{XY}^{AB} = C$ for all $X,Y,A,B$, so that $C^{RR}_{RL+LR}=2C$ and $C_{RL}^{RL}=C$. Furthermore the first order correction to $C_{XY}^{AB} = C$  vanishes
in the expansion. Therefore in the leading order of the non-relativistic expansion, we have
\beqa
\frac{\langle 0 \vert (B_{II})^{fg+gf}_{12} \vert 2{\rm N}, E\rangle }
 {\langle 0 \vert (B_{VI})^{fg+gf}_{12} \vert 2{\rm N}, E\rangle } &\simeq& 
 2 +  \rmO\left(\frac{{\vec p}^2}{m_Q^2}\right)
 \eeqa
for $(\sigma_1,\sigma_2)=(+,-)$ ( $^1S_0$ ), and
\beqa
\frac{\langle 0 \vert (B_{II})^{fg-fg}_{11} \vert 2{\rm N}, E\rangle }
 {\langle 0 \vert (B_{VI})^{fg-fg}_{11} \vert 2{\rm N}, E\rangle } &\simeq& 
 2 +  \rmO\left(\frac{{\vec p}^2}{m_Q^2}\right)
 \eeqa
for $(\sigma_1,\sigma_2)=(+,+)$ ( $^3S_1$ ). For both cases, we have  
positive sign for the ratio, which gives repulsion at short distance, 
the repulsive core.


\section{Conclusions and discussion}
The OPE analysis leads to conclusion that the S--state potential
at short distance behaves as in (\ref{summary1}) with (\ref{summary2}).
However perturbative considerations alone can not tell the 
crucial sign of the overall coefficient $C_E$.
Moreover we found that the latter was also not
directly predicted by chiral PT .
A crude estimation using non-relativistic quarks  
suggests that $C_E$ is positive, hence predicting a repulsive core, 
which diverges a little weaker than $r^{-2}$ at small $r$.
The leading corrections involve small powers of logs and hence 
it could happen that the dominant asymptotic behavior appears 
only at extremely short distances. 

Our analysis suggests that the repulsion of the NN potential
at short distance is related to the difference of anomalous 
dimensions between a 6--quark operator
and two 3--quark operators at 1-loop, and to the structure 
of the composite operators which probe the NN states.
The explicit 1--loop calculation indicates that a combination
of fermi statistics for quarks and the particular structure of the 
one gluon exchange interaction determines the sign and size of 
the $\beta$'s. The appearance of zero effective gamma eigenvalues 
is simply explained by chiral symmetry, however we were unable
to find a simple proof of the absence of positive eigenvalues
established by explicit calculation.
One speculation is that the latter intriguing pattern can be explained 
by the relation between the 1-loop QCD anomalous dimensions and those of
super YM theory \cite{Beisert}. 

At higher order in the perturbative expansion, tensor operators appear
in the OPE. Using this fact, we also found that
the tensor potential also diverges a little weaker than $r^{-2}$ as 
$r\rightarrow 0$.

There are several interesting extensions of the analysis using the OPE.
An application to the 3--flavor case may reveal
the nature of the repulsive core in the baryon-baryon potentials.
Since quark masses can be neglected in our OPE analysis,
the calculation can be done in the exact SU(3) symmetric limit.
It is also interesting to investigate the existence or the absence
of the repulsive core in the 3--body nucleon potential.
Such an investigation would require the  calculation
of anomalous dimensions of 9--quark operators at 2--loop level.
Certainly more precise evaluations (also involving numerical simulations)
of matrix elements $\langle 0 \vert O_X\vert E\rangle$ 
will also be needed to theoretically predict 
the nature of the core of the NN potential.

\section*{Acknowledgments}
S.~A would like to thank Dr. T.~Doi, Prof. T.~Hatsuda, Dr. N.~Ishii and Prof. W.~Wise for useful discussions.
S.~A. is supported in part by Grant-in-Aid of the Ministry of Education, 
Sciences and Technology, Sports and Culture (Nos. 20340047, 20105001, 20105003).
S.~A. and J.~B. would like to thank the Max-Planck-Institut f\"ur Physik 
for its kind hospitality during our stay for this research project.

\appendix

\section{Explicit calculations of the divergent part for 6--quark operators at 1-loop}
\label{appendixA}

In determining the divergent parts for 6--quark operators we will 
consider the various cases in turn adopting the following mechanical (if 
rather inelegant) procedure. We first list the various operators 
which can appear and determine their linear relations. Then we compute the
divergent parts using (\ref{eq:T0}), initially keeping $N$ explicit in the 
formulae to indicate from which part of (\ref{eq:T0}) the terms originate.
Finally we set $N=3$ and use the constraint equations to express the 
result in terms of linearly independent operators.
 
\subsection{$S=0$ and $I=1$ case}
{\bf 1.} $B_I^{01}=B_{\alpha[\beta\alpha]}^{ffg}B_{\beta[\beta\alpha]}^{ffg}+
B_{\hat\alpha[\hat\beta\hat\alpha]}^{ffg}B_{\hat\beta[\hat\beta\hat\alpha]}^{ffg}$\\
\\
\noindent
We prepare following 11 operators
\beqa
B_1 &=& B^{ffg}_{\alpha\alpha\beta}B^{ffg}_{\alpha\beta\beta}, \
B_2 = B^{ffg}_{\alpha\beta\alpha}B^{ffg}_{\beta\beta\alpha}, \
B_3= B^{ffg}_{\alpha\beta\alpha}B^{ffg}_{\alpha\beta\beta}, \
B_4= B^{ffg}_{\alpha\alpha\beta}B^{ffg}_{\beta\beta\alpha}, \nn \\
B_5&=& B^{ffg}_{\alpha\alpha\alpha}B^{ffg}_{\beta\beta\beta}, \
B_6 = B^{fff}_{\alpha\alpha\beta}B^{fgg}_{\alpha\beta\beta}, \
B_7 = B^{fff}_{\alpha\alpha\beta}B^{fgg}_{\beta\alpha\beta}, \
B_8 = B^{fff}_{\alpha\alpha\alpha}B^{fgg}_{\beta\beta\beta}, \nn \\
B_9&=& B^{fff}_{\beta\beta\beta}B^{fgg}_{\alpha\alpha\alpha},\
B_{10} = B^{fff}_{\alpha\beta\beta}B^{fgg}_{\alpha\alpha\beta}, \
B_{11} = B^{fff}_{\alpha\beta\beta}B^{fgg}_{\beta\alpha\alpha}\,,
\eeqa
which are all the possible operators corresponding to the Dirac
labels 111222 in Table~\ref{T3f3ga} (in the $4f2g$ case). 
In terms of these we write
\beqa
B_{\alpha[\beta\alpha]}^{ffg}B_{\beta[\beta,\alpha]}^{ffg} &=&
B_1+B_2-B_3-B_4\,.
\eeqa
Taking $B_{1,2,3,4}$ as independent operators, the constraints from 
(\ref{eq:constraint}) are:
\beqa
B_5 &=& B_4 - 4B_3, \quad
B_6 = B_1, \quad B_7 = 2 B_3 - B_4, \quad B_8 = - 3 B_1 \nn \\
B_9 &=& 3 B_2, \quad B_{10} = B_4 - 2 B_3, \quad B_{11} = - B_2\,.
\label{gi1}
\eeqa
The 1-loop corrections $\Gamma^{(1)}$ to $B_{1,2,3,4}$ can be calculated from 
${\bf T}_0$ as
\beqa
\Gamma^{(1)}_1 &=& (5-9N) B_1 -4NB_3-2(N+1)B_4-2B_5
+N(-3B_6 -2 B_7 + B_8 + 4 B_{10}),\nn \\
\Gamma^{(1)}_2 &=&(5-9N) B_2 -4NB_3-2(N+1)B_4-2B_5
+N(2B_{10}+3B_{11}-4B_7-B_9)\,,\nn \\
\Gamma^{(1)}_3 &=&(1-11N)B_3 -2N(B_1+B_2) +(2-N)(B_4-B_5)+2N(B_{10}+B_{11}
-B_6-B_7)\,,\nn \\
\Gamma^{(1)}_4 &=&(3-8N)B_4 -4(N+1)(B_1+B_2)
+4(2-N)B_3 +(N-2)B_5 +2N(B_{10}-B_7)\,,\nn \\
\eeqa
where an overall factor $d/\epsilon$ is dropped on the rhs for simplicity.
i.e. using the gauge identities eq.~(\ref{gi1}) we obtain  
\begin{equation}
\Gamma^{(1)}_i=\frac{1}{2\epsilon}\gamma_{ij}B_j\,,
\end{equation}
with the matrix $\gamma$ given by
\beqa
\gamma/(2d) &=& \left(\begin{array}{cccc}
-40 &   0 & -40 &  8 \\
  0 & -40 & -40 &  8 \\
-12 & -12 & -60 & 12 \\
-16 & -16 & -32 & -8 \\
\end{array}
\right)\,. 
\eeqa
So we obtain
\beqa
\Gamma^{(1)}_{1+2-3-4} &=& -12\frac{d}{\epsilon}B_{1+2-3-4}\,.
\eeqa
For $B_{\hat\alpha[\hat\beta,\hat\alpha]}^{ffg}B_{\hat\beta[\hat\beta,\hat\alpha]}^{ffg}$, 
the result can be obtained from the above by the interchange of $\alpha$ and $\hat \alpha$. 
In terms of $\tilde B_i = B_i(\alpha\leftrightarrow \hat\alpha)$, we have
\beqa
\left(\Gamma_I^{01}\right)^{(1)}&\equiv&
\left[ (B+\tilde B)_{1+2-3-4}\right]^{\rm 1-loop,div} 
=-12\frac{d}{\epsilon} (B +\tilde B)_{1+2-3-4}\,.
\eeqa
This corresponds to the result for the first $I=1$ operator 
for the entry 111222 in Table~\ref{T3f3ga}. 
The other eigenvalues of $\gamma$ are obtained similarly. 

\noindent
{\bf 2.} $B_{II}^{01}=B^{ffg}_{\alpha[\beta\alpha]}B^{ffg}_{\beta[\hat\beta\hat\alpha]} + B^{ffg}_{\alpha[\hat\beta\hat\alpha]}B^{ffg}_{\beta[\beta\alpha]} + (\alpha,\beta \leftrightarrow \hat\alpha,\hat\beta) $\\
\\
\noindent
We prepare the following 6 operators,
\beqa
B_1 &=& B^{ffg}_{\alpha\beta\alpha}B^{ffg}_{\beta[\hat\beta\hat\alpha]}, \ 
B_2 = B^{ffg}_{\alpha\alpha\beta}B^{ffg}_{\beta[\hat\beta\hat\alpha]}, \
B_3 = B^{ffg}_{\beta\beta\alpha}B^{ffg}_{\alpha[\hat\beta\hat\alpha]} \nn \\
B_4&=& B^{ffg}_{\alpha\beta\beta}B^{ffg}_{\alpha[\hat\beta\hat\alpha]}, \
B_5 = B^{fff}_{\alpha\beta\beta}B^{gfg}_{\alpha[\hat\beta\hat\alpha]}, \ 
B_6 = B^{fff}_{\alpha\alpha\beta}B^{gfg}_{\beta[\hat\beta\hat\alpha]} ,
\eeqa
in terms of which we have
\beqa
B^{ffg}_{\alpha[\beta\alpha]}B^{ffg}_{\beta[\hat\beta\hat\alpha]} + B^{ffg}_{\alpha[\hat\beta\hat\alpha]}B^{ffg}_{\beta[\beta\alpha]}
&=& B_1-B_2 -B_3 + B_4\,.
\eeqa
The 1-loop corrections to $B_{1,2,3,4}$ can be calculated as
\beqa
\Gamma^{(1)}_1 &=& (3-N) B_1-2(N+1)B_2+(N-2)B_3-2B_4+N B_5  -2N B_6\,, \nn \\ 
\Gamma^{(1)}_2 &=& 4 B_2-4(N+1)B_1 +2(N-2)B_4-NB_6\,, \nn \\
\Gamma^{(1)}_3 &=& 4 B_3-4(N+1)B_4 +2(N-2)B_1-NB_5\,, \nn \\
\Gamma^{(1)}_4 &=& (3-N) B_4-2(N+1)B_3 +(N-2)B_2-2B_1+N B_6  -2N B_5\,.
\eeqa
Therefore we have
\beqa
\Gamma^{(1)}_{1-2-3+4}&=&12\frac{d}{\epsilon}B_{1-2-3+4}\,.
\eeqa
For $B^{ffg}_{\hat\alpha[\hat\beta\hat\alpha]}B^{ffg}_{\hat\beta[\beta\alpha]} + B^{ffg}_{\hat\alpha[\beta\alpha]}B^{ffg}_{\hat\beta[\hat\beta\hat\alpha]}$,
we introduce $\tilde B_i = B_i(\alpha\leftrightarrow \hat\alpha)$, so that we have
\beqa
\left(\Gamma_{II}^{01}\right)^{(1)}&\equiv&\left[ (B+\tilde B)_{1-2-3+4}\right]^{\rm 1-loop,div}
=12\frac{d}{\epsilon} (B +\tilde B)_{1-2-3+4}\,.
\eeqa

\noindent 
{\bf 3.} $B_{III}^{01}=B^{ffg}_{\alpha[\beta\alpha]} B^{ffg}_{\hat\beta[\beta\alpha]} +
B^{ffg}_{\hat\alpha[\beta\alpha]} B^{ffg}_{\beta[\beta\alpha]}
+  (\alpha,\beta \leftrightarrow \hat\alpha,\hat\beta) $\\
\\ \noindent
We prepare the following 15 operators:
\beqa
B_1 &=& B^{ffg}_{\alpha\beta\alpha}B^{ffg}_{\hat\beta\beta\alpha} , \
B_2 = B^{ffg}_{\alpha\beta\alpha}B^{ffg}_{\hat\beta\alpha\beta}, \
B_3 = B^{ffg}_{\alpha\alpha\beta}B^{ffg}_{\hat\beta\beta\alpha}, \
B_4 = B^{ffg}_{\alpha\alpha\beta}B^{ffg}_{\hat\beta\alpha\beta},  \
B_5 = B^{ffg}_{\beta\beta\alpha}B^{ffg}_{\hat\beta\alpha\alpha}, \nn \\
B_6 &=& B^{ffg}_{\alpha\beta\beta}B^{ffg}_{\hat\beta\alpha\alpha} , \
B_7 = B^{ffg}_{\alpha\alpha\alpha}B^{ffg}_{\hat\beta\beta\beta} , \
C_1 = B^{fff}_{\alpha\beta\beta}B^{fgg}_{\hat\beta\alpha\alpha}, \
C_2 = B^{fff}_{\alpha\alpha\beta}B^{fgg}_{\hat\beta\alpha\beta}, \ 
C_3 = B^{fff}_{\alpha\alpha\alpha}B^{fgg}_{\hat\beta\beta\beta} , \nn\\
D_1 &=& B^{fgg}_{\alpha\alpha\beta}B^{fff}_{\hat\beta\alpha\beta}, \
D_2 = B^{fgg}_{\beta\alpha\alpha}B^{fff}_{\hat\beta\alpha\beta} . \ 
D_3 = B^{fgg}_{\alpha\beta\beta}B^{fff}_{\hat\beta\alpha\alpha} , \
D_4 = B^{fgg}_{\beta\alpha\beta}B^{fff}_{\hat\beta\alpha\alpha}, \
D_5 = B^{fgg}_{\alpha\alpha\alpha}B^{fff}_{\hat\beta\beta\beta}, \nn
\eeqa
in terms of which we write
\beqa
B^{ffg}_{\alpha[\beta\alpha]} B^{ffg}_{\hat\beta[\beta\alpha]} &=& 
B_1-B_2-B_3+B_4\,.
\eeqa
Taking $B_{1,2,3,4,5}$ as independent operators, others can be expressed as
\beqa
B_6 &=& - B_2, \ B_7=B_3-4B_2,  \nn \\
C_1&=&-2B_1-B_5,\ C_2=D_1=-D_4 =2B_2-B_3, \ C_3= - 3B_4, \nn \\
D_2&=& - B_5, D_3 = - B_4,\ D_5 = B_5 - 2B_1.
\eeqa
The 1-loop corrections to $B_{1,2,3,4}$ can be calculated as
\beqa
\Gamma^{(1)}_1 &=& -4N B_1-2(N+1)(B_2+B_3)
+(N-2) B_5 -2(B_6+B_7)\nn\\ 
&+& N(C_1-2C_2-2D_1-D_2+D_5)\nn \\
&=&-24 B_1-22 B_2 + 2 B_3 + 4B_5\,, \nn \\
\Gamma^{(1)}_2 &=& -5N B_2-2(N+1)(B_1+B_4)
- 2B_5 +(N-2)(B_6+B_7)\nn \\ 
&+&N(D_4-C_2-D_1-2D_2)\nn \\
&=&-8B_1-38 B_2 + 10 B_3 -8B_4+ 4 B_5\,, \nn \\
\Gamma^{(1)}_3 &=& (2-4N)B_3 -2(N+1)(B_4+2B_1)
+ (N-2)(2B_6+ B_7) - N(C_2+ 2D_1) \nn \\
&=& -16B_1-24B_2 -8B_4\,,\nn \\
\Gamma^{(1)}_4 &=& (2- N) B_4 -2(N+1)(B_3+2B_2)
-2(2B_6+B_7) +N(C_3-2C_2+2D_3-4D_1)\nn \\
&=& -40B_2+8 B_3 - 16 B_4\,.
\eeqa
We therefore obtain
\beqa
\Gamma^{(1)}_{1-2-3+4}&=& 0\,.
\eeqa
We have 3 more structures:
$B^{ffg}_{\hat\alpha[\beta\alpha]} B^{ffg}_{\beta[\beta\alpha]}$,
$B^{ffg}_{\hat\alpha[\hat\beta\hat\alpha]} B^{ffg}_{\beta[\hat\beta\hat\alpha]} $ and
$B^{ffg}_{\alpha[\hat\beta\hat\alpha]} B^{ffg}_{\hat\beta[\hat\beta\hat\alpha]} $.
We introduce operators $B^\prime_i = - B_i(\alpha\leftrightarrow\beta)$ 
for the first one,
$\tilde B_i = B_i(\alpha\leftrightarrow\hat\alpha)$ for the second, and
$\hat B_i = B^\prime_i(\alpha\leftrightarrow\hat\alpha)$ for the third. 
We then have
\beqa
\left(\Gamma_{III}^{01}\right)^{(1)}&=&0\,.
\eeqa 

\noindent
{\bf 4.} $B_{IV}^{01}=B^{ffg}_{\alpha[\hat\beta\hat\alpha]} B^{ffg}_{\beta[\hat\beta\hat\alpha]}
+B^{ffg}_{\hat\alpha[\beta\alpha]} B^{ffg}_{\hat\beta[\beta\alpha]}$\\
\\ \noindent
We prepare the following 12 operators:
\beqa
B_1 &=& B^{ffg}_{\alpha\hat\beta\hat\alpha}B^{ffg}_{\beta\hat\beta\hat\alpha} , \
B_2 = B^{ffg}_{\alpha\hat\beta\hat\alpha}B^{ffg}_{\beta\hat\alpha\hat\beta}, \
B_3 = B^{ffg}_{\alpha\hat\alpha\hat\beta}B^{ffg}_{\beta\hat\beta\hat\alpha}, \
B_4 = B^{ffg}_{\alpha\hat\alpha\hat\beta}B^{ffg}_{\beta\hat\alpha\hat\beta}, \
B_5 = B^{ffg}_{\alpha\hat\alpha\hat\alpha}B^{ffg}_{\beta\hat\beta\hat\beta} ,  \nn \\
B_6 &=& B^{ffg}_{\alpha\hat\beta\hat\beta}B^{ffg}_{\beta\hat\alpha\hat\alpha} , \
C_1 = B^{fff}_{\alpha\hat\beta\hat\beta}B^{fgg}_{\beta\hat\alpha\hat\alpha}, \
C_2 = B^{fff}_{\alpha\hat\alpha\hat\alpha}B^{fgg}_{\beta\hat\beta\hat\beta} , \
C_3 = B^{fff}_{\alpha\hat\alpha\hat\beta}B^{fgg}_{\beta\hat\alpha\hat\beta}, \
D_1 = B^{fgg}_{\alpha\hat\beta\hat\beta}B^{fff}_{\beta\hat\alpha\hat\alpha}, \nn \\
D_2 &=& B^{fgg}_{\alpha\hat\alpha\hat\alpha}B^{fff}_{\beta\hat\beta\hat\beta} , \
D_3 = B^{fgg}_{\alpha\hat\alpha\hat\beta}B^{fff}_{\beta\hat\alpha\hat\beta}, \nn
\eeqa
in terms of which we write
\beqa
B^{ffg}_{\alpha[\hat\beta\hat\alpha]} B^{ffg}_{\beta[\hat\beta\hat\alpha]} 
&=& B_1-B_2-B_3 +B_4\,.
\eeqa
The constraint from gauge invariance leads to
\beqa
4 B_1 + C_1+ D_2 &=& 0, \
4B_4 + C_2 + D_1 = 0, \
B_2+B_3 + B_5 + B_6 + C_3+ D_3 = 0.
\eeqa
The 1-loop corrections can be calculated as
\beqa
\Gamma^{(1)}_1 &=& (5-3N) B_1-2(N+1)(B_2+B_3)-2(B_5+B_6)+N(D_2-2D_3+C_1-2C_3)\,, \nn \\
\Gamma^{(1)}_2 &=& (3-4N) B_2-2(N+1)(B_1+B_4)+(N-2)(B_5+B_6-B_3)-N(C_3+D_3)\,, \nn \\
\Gamma^{(1)}_3 &=& (3-4N) B_3-2(N+1)(B_1+B_4)+(N-2)(B_5+B_6-B_2)-N(C_3+D_3)\,,\nn \\
\Gamma^{(1)}_4 &=& (5-3N) B_4-2(N+1)(B_2+B_3)-2(B_5+B_6)+N(D_1-2D_3+C_2-2C_3)\nn\,.\\ 
&&
\eeqa
Therefore we have
\beqa
\Gamma^{(1)}_{1-2-3+4}  &=& 0\,.
\eeqa
We introduce $\tilde B_i = B_i(\alpha\leftrightarrow\hat\alpha)$ 
for $B^{ffg}_{\hat \alpha[\beta\alpha]} B^{ffg}_{\hat\beta[\beta\alpha]} $, 
so that we have
\beqa
\left(\Gamma_{IV}^{01}\right)^{(1)}&=&0\,.
\eeqa

\noindent
{\bf 5.}  $B_{V}^{01}=B^{ffg}_{\alpha[\hat\beta\hat\alpha]} B^{ffg}_{\hat\beta[\beta\alpha]} 
B^{ffg}_{\hat\alpha[\beta\alpha]} B^{ffg}_{\beta[\hat\beta\hat\alpha]} $
and $B_{VI}^{01}=B^{ffg}_{\alpha[\beta\alpha]} B^{ffg}_{\hat\beta[\hat\beta\hat\alpha]} 
B^{ffg}_{\hat\alpha[\hat\beta\hat\alpha]} B^{ffg}_{\beta[\beta\alpha]} $\\
\\ \noindent
In this case we have to prepare 29 operators as follows:
\beqa
B_1 &=& B^{ffg}_{\alpha\hat\beta\hat\alpha} B^{ffg}_{\hat\beta\beta\alpha}, \
B_2 = B^{ffg}_{\alpha\hat\beta\hat\alpha} B^{ffg}_{\hat\beta\alpha\beta}, \
B_3 = B^{ffg}_{\alpha\hat\alpha\hat\beta} B^{ffg}_{\hat\beta\beta\alpha}, \
B_4 = B^{ffg}_{\alpha\hat\alpha\hat\beta} B^{ffg}_{\hat\beta\alpha\beta}, \
B_5 = B^{ffg}_{\beta\hat\beta\hat\alpha} B^{ffg}_{\hat\beta\alpha\alpha}, \nn \\
B_6 &=& B^{ffg}_{\beta\hat\alpha\hat\beta} B^{ffg}_{\hat\beta\alpha\alpha}, \
B_7 = B^{ffg}_{\alpha\hat\beta\hat\beta} B^{ffg}_{\hat\alpha\beta\alpha}, \
B_8 = B^{ffg}_{\alpha\hat\beta\hat\beta} B^{ffg}_{\hat\alpha\alpha\beta}, \
C_1 = B^{fff}_{\alpha\hat\beta\hat\beta} B^{fgg}_{\beta\hat\alpha\alpha}, \ 
C_2 = B^{fff}_{\alpha\hat\beta\hat\beta} B^{fgg}_{\alpha\hat\alpha\beta}, \nn \\ 
C_3 &=& B^{fff}_{\alpha\hat\alpha\hat\beta} B^{fgg}_{\beta\hat\beta\alpha}, \
C_4 = B^{fff}_{\alpha\hat\alpha\hat\beta} B^{fgg}_{\alpha\hat\beta\beta}, \
D_1= B^{fgg}_{\hat\beta\alpha\hat\alpha} B^{fff}_{\hat\beta\alpha\beta}, \  
D_2= B^{fgg}_{\hat\beta\beta\hat\alpha} B^{fff}_{\hat\beta\alpha\alpha}, \  
D_3= B^{fgg}_{\hat\alpha\alpha\hat\beta} B^{fff}_{\hat\beta\alpha\beta}, \nn \\ 
D_4&=& B^{fgg}_{\hat\alpha\beta\hat\beta} B^{fff}_{\hat\beta\alpha\alpha}, \
X_1 = B^{ffg}_{\hat\beta\hat\beta\hat\alpha} B^{ffg}_{\alpha\beta\alpha}, \
X_2 = B^{ffg}_{\alpha\beta\hat\alpha} B^{ffg}_{\hat\beta\hat\beta\alpha}, \
X_3 = B^{fgg}_{\alpha\alpha\hat\alpha} B^{fff}_{\hat\beta\hat\beta\beta}, \
Y_1 = B^{ffg}_{\hat\beta\hat\beta\hat\alpha} B^{ffg}_{\alpha\alpha\beta}, \nn \\
Y_2 &=& B^{ffg}_{\alpha\alpha\hat\alpha} B^{ffg}_{\hat\beta\hat\beta\beta}, \
Z_1 =B^{ffg}_{\hat\alpha\hat\beta\hat\beta} B^{ffg}_{\alpha\beta\alpha}, \
Z_2 =B^{ffg}_{\alpha\beta\hat\beta} B^{ffg}_{\hat\alpha\hat\beta\alpha}, \
Z_3 =B^{ffg}_{\alpha\hat\alpha\alpha} B^{ffg}_{\hat\beta\beta\hat\beta}, \
Z_4 =B^{fff}_{\alpha\hat\alpha\beta} B^{fgg}_{\hat\beta\hat\beta\alpha}, \nn \\
Z_5 &=& B^{fgg}_{\alpha\alpha\hat\beta} B^{fff}_{\hat\beta\beta\hat\alpha}, \
V_1 = B^{ffg}_{\hat\alpha\hat\beta\hat\beta} B^{ffg}_{\alpha\alpha\beta}, \
V_2 = B^{ffg}_{\alpha\alpha\hat\beta} B^{ffg}_{\hat\alpha\hat\beta\beta}, \
V_3 = B^{fff}_{\alpha\hat\alpha\alpha} B^{fgg}_{\hat\beta\hat\beta\beta}, \nn
\eeqa
in terms of which we have
\beqa
B^{ffg}_{\alpha[\hat\beta\hat\alpha]} B^{ffg}_{\hat\beta[\beta\alpha]}
&=& B_1- B_2 - B_3 +B_4 \equiv{\cal B}_5 , \\
B^{ffg}_{\alpha [\beta,\alpha]}B^{ffg}_{\hat\beta [\hat\beta,\hat\alpha]}
&=&-X_1+Y_1+Z_1-V_1 \equiv {\cal B}_6\,.
\eeqa
The constraint from gauge invariance gives
\beqa
C_1 &=& -2B_1 - X_1,\ D_1=-B_1-B_5-X_1, \
X_2=-2(B_1+B_5)-X_1,\
 X_3 = 2B_5+X_1, \nn \\
D_2 &=& C_2 =-2B_2-Y_1,\ Y_2 = - 4B_2-Y_1, \
Z_3=Z_2+B_6-C_3,\ Z_4=D_3+Z_1-Z_2, \nn \\
Z_5&=&B_6+Z_1-Z_3, \
B_6+D_3 =B_7+C_3,\ B_3+Z_1+B_6+D_3 = 0 , \nn \\
C_4 &=& -B_4-B_8-V_1, \
D_4= -2B_4-V_1,\
 V_2=-2B_4-2B_8-V_1,\
V_3=2B_8+V_1,
\eeqa
where $B_{1,2,3,4,5,6,7,8}$,  $X_1$, $Y_1$, $Z_{1,2}$ and $V_1$ 
are taken to be independent.
The 1-loop corrections to $B_{1,2,3,4}$  can be calculated as
\beqa
\Gamma^{(1)}_1 &=& (2-N)B_1-2(N+1)(B_2+B_3)+(N-2)B_5 - 2B_7 +N(C_1-2C_3-D_1)\,,\nn  \\
\Gamma^{(1)}_2 &=&2B_2-2(N+1)(B_1+B_4)-2(B_5+B_8)+N(C_2-2C_4+D_2-2D_1)\,,\nn \\
\Gamma^{(1)}_3 &=& (2-2N) B_3 -2(N+1)(B_1+B_4)+(N-2)(B_6+B_7)- N(C_3+D_3)\,, \nn \\
\Gamma^{(1)}_4 &=& (2-N)B_4-2(N+1)(B_2+B_3)-2B_6+(N-2)B_8 +N(-C_4-2D_3+D_4)\,,\nn\\
&&
\eeqa
which leads to
\beqa
\Gamma_{1-2-3+4}^{(1)} &=& 3(N+2)(B_1+B_4) -(4N+6)B_2 -(2N+6) B_3
+N(B_5-B_6 -B_7+B_8)\nn \\
& +& N(C_1-C_2-C_3+C_4+D_1-D_2-D_3+D_4)\,.
\eeqa
On the other hand, constraints give
\beqa
C_1-C_2-C_3+C_4 &+& D_1-D_2-D_3+D_4 =
-3B_1 +4B_2 + 2B_3 -3B_4
\nn \\
&-&B_5+B_6+B_7-B_8 +2(-X_1+Y_1+Z_1-V_1)\,,
\eeqa
so we finally obtain
\beqa
\Gamma_{1-2-3+4}^{(1)}  &=& 6\frac{d}{\epsilon}{\cal B}_5 +  
6\frac{d}{\epsilon}{\cal B}_6\,.
\eeqa
We introduce operators 
$\tilde B_i = B_i(\alpha\leftrightarrow\hat\alpha)$ for ${\cal B}_{5,6}$, so that we have
\beqa
\left(\Gamma_{V}^{01}\right)^{(1)}&=& 
6\frac{d}{\epsilon}({\cal B} +\tilde{\cal  B})_5
+6\frac{d}{\epsilon}({\cal B} +\tilde{\cal  B})_6\,, \quad
\left(\Gamma_{VI}^{01}\right)^{(1)}
=24\frac{d}{\epsilon}({\cal B} +\tilde{\cal  B})_6\,. 
\eeqa

\subsection{ $S=1$ and $I=0$ case}
{\bf 1.} $B_I^{10}=B^{ffg}_{\alpha [\beta,\alpha]}B^{ggf}_{\alpha [\beta,\alpha]}
+B^{ffg}_{\hat\alpha [\hat\beta,\hat\alpha]}B^{ggf}_{\hat\alpha [\hat\beta,\hat\alpha]}$\\
\\ \noindent
We prepare the following 11 operators:
\beqa
B_1 &=& B^{ffg}_{\alpha\beta\alpha}B^{ggf}_{\alpha\beta\alpha}, \
B_2 = B^{ffg}_{\alpha\beta\alpha}B^{ggf}_{\alpha\alpha\beta}, \
B_3= B^{ffg}_{\alpha\alpha\beta}B^{ggf}_{\alpha\beta\alpha}, \
B_4= B^{ffg}_{\alpha\alpha\beta}B^{ggf}_{\alpha\alpha\beta},  \
C_1= B^{ffg}_{\alpha\beta\beta}B^{ggf}_{\alpha\alpha\alpha}, \nn \\
C_2 &=& B^{ffg}_{\alpha\alpha\alpha}B^{ggf}_{\alpha\beta\beta}, \
C_3 = B^{ffg}_{\beta\beta\alpha}B^{ggf}_{\alpha\alpha\alpha}, \
C_4 = B^{ffg}_{\alpha\alpha\alpha}B^{ggf}_{\beta\beta\alpha}, \
C_5= B^{fff}_{\alpha\alpha\beta}B^{ggg}_{\alpha\alpha\beta},\
C_6 = B^{fff}_{\alpha\beta\beta}B^{ggg}_{\alpha\alpha\alpha}, \nn \\
C_7 &=& B^{fff}_{\alpha\alpha\alpha}B^{ggg}_{\alpha\beta\beta}, \nn
\eeqa
in terms of which
\beqa
B^{ffg}_{\alpha [\beta,\alpha]}B^{ggf}_{\alpha [\beta,\alpha]} 
&=& B_1-B_2-B_3+B_4\,.
\eeqa
Constraints are given by
\beqa
C_1&=& C_2 = 2B_1 - B_4, \ C_5 = B_4 - 4B_1, \
C_3= B_2, \ C_6 = - 3 B_2 , \
C_4= B_3,\  C_7 = - 3 B_3 .  \nn
\eeqa
The 1-loop corrections $\Gamma^{(1)}$ can be calculated as
\beqa
\Gamma^{(1)}_1 &=& -(3+7N) B_1 - 2(B_2+B_3) +N(B_4-C_5) 
-2(C_1+C_2+C_3+C_4)\,, \nn \\
\Gamma^{(1)}_2 &=& -(1+3N) B_2 - 4B_1 -2(N+1)B_4-2C_1-4C_2-(N+2)C_3
+N(C_6-2C_5)\,,\nn \\
\Gamma^{(1)}_3 &=& -(1+3N) B_3 - 4B_1 -2(N+1)B_4-4C_1-2C_2-(N+2)C_4
+N(C_7-2C_5)\,,\nn \\
\Gamma^{(1)}_4 &=& (1-6N) B_4 -4(N+1)(B_2+B_3)+ 4N B_1 +2(N-2)(C_1+C_2)-NC_5\,,
\eeqa
and therefore we have
\beqa
\Gamma_{1-2-3+4}^{(1)} &=& -4\frac{d}{\epsilon}B_{1-2-3+4}\,.
\eeqa
For $B^{ffg}_{\hat\alpha [\hat\beta,\hat\alpha]}B^{ggf}_{\hat\alpha [\hat\beta,\hat\alpha]} $, the result can be obtained from the above by the interchange of $\alpha$ and $\hat\alpha$. In terms of $\tilde B_i = B_i(\alpha\leftrightarrow\hat\alpha)$, we have
\beqa
\left(\Gamma_{I}^{10}\right)^{(1)}&=&
-4\frac{d}{\epsilon} (B+\tilde B)_{1-2-3+4}\,. 
\eeqa

\noindent
{\bf 2.} $B_{II}^{10}=B^{ffg}_{\alpha[\beta\alpha]}B^{ggf}_{\alpha[\hat\beta\hat\alpha]} +
B^{ffg}_{\hat\alpha[\hat\beta\hat\alpha]}B^{ggf}_{\hat\alpha[\beta\alpha]} 
 +(\alpha,\beta \leftrightarrow \hat\alpha,\hat\beta) $ \\
 \\ \noindent
 We prepare the following 6 operators:
\beqa
B_1 &=& B^{ffg}_{\alpha\beta\alpha}B^{ggf}_{\alpha[\hat\beta\hat\alpha]}, \
B_2 = B^{ffg}_{\alpha\alpha\beta}B^{ggf}_{\alpha[\hat\beta\hat\alpha]}, \
B_3 = B^{ffg}_{\alpha[\hat\beta\hat\alpha]}B^{ggf}_{\alpha\beta\alpha}, \
B_4 = B^{ffg}_{\alpha[\hat\beta\hat\alpha]}B^{ggf}_{\alpha\alpha\beta}, \nn \\
C_1 &=& B^{ffg}_{\alpha\alpha\alpha}B^{ggf}_{\beta[\hat\beta\hat\alpha]}, \
C_2 = B^{ffg}_{\beta[\hat\beta\hat\alpha]}B^{ggf}_{\alpha\alpha\alpha}, \nn
\eeqa
in terms of which
\beqa
B^{ffg}_{\alpha[\beta\alpha]}B^{ggf}_{\alpha[\hat\beta\hat\alpha]} +
B^{ffg}_{\hat\alpha[\hat\beta\hat\alpha]}B^{ggf}_{\hat\alpha[\beta\alpha]}
&=&
B_1 - B_2 +B_3 -B_4\,.
\eeqa
The 1-loop corrections can be calculated as
\beqa
\Gamma^{(1)}_1 &=& (N+1)(B_1-2B_2) -N(2B_3+B_4) +NC_2 -2 C_1\,, \nn \\
\Gamma^{(1)}_2 &=& (2N+3)B_2-4(N+1)B_1 -2NB_3 +(N-2)C_1\,,  \nn \\
\Gamma^{(1)}_3 &=& (N+1)(B_3-2B_4) -N(2B_1+B_2) +NC_1 -2 C_2\,,  \nn \\
\Gamma^{(1)}_4 &=& (2N+3)B_4-4(N+1)B_3 -2NB_1 +(N-2)C_2\,,
\eeqa
and therefore we have
\beqa
\Gamma^{(1)}_{1-2+3-4} &=& 20\frac{d}{\epsilon}B_{1-2+3-4}\,. 
\eeqa
For $B^{ffg}_{\hat\alpha[\hat\beta\hat\alpha]}B^{ggf}_{\hat\alpha[\beta\alpha]} +
B^{ffg}_{\alpha[\beta\alpha]}B^{ggf}_{\alpha[\hat\beta\hat\alpha]} $, 
we introduce $\tilde B_i = B_i(\alpha\leftrightarrow\hat\alpha)$, so that we have
\beqa
\left(\Gamma_{II}^{10}\right)^{(1)}&=&
20\frac{d}{\epsilon} (B+\tilde B)_{1-2+3-4}\,. 
\eeqa

\noindent
{\bf 3.} $B_{III}^{10}=B^{ffg}_{\alpha[\beta\alpha]}B^{ggf}_{\hat\alpha[\beta\alpha]}
+ B^{ffg}_{\hat\alpha[\beta\alpha]}B^{ggf}_{\alpha[\beta\alpha]}
+(\alpha,\beta \leftrightarrow \hat\alpha,\hat\beta) $\\
 \\ \noindent
We prepare the following 15 operators:
\beqa
B_1 &=& B^{ffg}_{\alpha\beta\alpha}B^{ggf}_{\hat\alpha\beta\alpha}, \
B_2 = B^{ffg}_{\alpha\beta\alpha}B^{ggf}_{\hat\alpha\alpha\beta}, \
B_3 = B^{ffg}_{\alpha\alpha\beta}B^{ggf}_{\hat\alpha\beta\alpha}, \
B_4 = B^{ffg}_{\alpha\alpha\beta}B^{ggf}_{\hat\alpha\alpha\beta}, \
C_1 = B^{ffg}_{\alpha\alpha\alpha}B^{ggf}_{\hat\alpha\beta\beta}, \nn \\
C_2 &=& B^{ffg}_{\alpha\beta\beta}B^{ggf}_{\hat\alpha\alpha\alpha}, \
C_3 = B^{ffg}_{\beta\beta\alpha}B^{ffg}_{\hat\alpha\alpha\alpha}, \
C_4 = B^{ggf}_{\alpha\beta\alpha}B^{ggf}_{\alpha\beta\hat\alpha}, \
C_5 = B^{ggf}_{\alpha\alpha\beta} B^{ffg}_{\alpha\beta\hat\alpha}, \
C_6 = B^{ggf}_{\alpha\beta\beta} B^{ffg}_{\alpha\alpha\hat\alpha}, \nn \\
C_7 &=& B^{fff}_{\alpha\alpha\beta} B^{ggg}_{\hat\alpha\alpha\beta}, \
C_8 = B^{fff}_{\alpha\beta\beta} B^{ggg}_{\hat\alpha\alpha\alpha}, \
C_9 = B^{ggf}_{\alpha\alpha\alpha} B^{ffg}_{\beta\beta\hat\alpha}, \
C_{10} = B^{fff}_{\alpha\alpha\alpha} B^{ggg}_{\hat\alpha\beta\beta}, \
C_{11} = B^{ggf}_{\beta\beta\alpha} B^{ffg}_{\alpha\alpha\hat\alpha}, \nn
\eeqa
in terms of which we write
\beqa
B^{ffg}_{\alpha[\beta\alpha]}B^{ggf}_{\hat\alpha[\beta\alpha]} 
&=& B_1-B_2 -B_3 + B_4\,.
\eeqa
Constraints are given by
\beqa
2C_2 &=& 2B_1-B_4+C_1, \ 2C_4= -C_1-B_4,\
 C_6=B_4-2B_1,\
C_7=-C_1-2B_1, \nn \\
C_5 &=& -C_3,\ C_8= -2B_2 -C_3,\ C_9 = -2B_2+C_3, \
C_{10}= - 3 B_3,\ C_{11} = - B_3. \nn
\eeqa
The 1-loop corrections can be calculated as
\beqa
\Gamma^{(1)}_1 &=& - 5N B_1-2(N+1)(B_2+B_3) 
+(N-2)(C_1+C_2)-2C_3\nn\\
&+&N(-C_4-2C_5+C_6-C_7)\,,\nn \\
\Gamma^{(1)}_2 &=& - 4N B_2 -2(N+1)(B_1+B_4)
-2(C_1+C_2)+(N-2)C_3\nn\\
&+&N(-2C_4-C_5-2C_7+C_8+C_9)\,,\nn \\
\Gamma^{(1)}_3 &=& (2- N)B_3-2(N+1)(B_4+2B_1) 
-2(C_1+2C_2)+N(-4C_4-2C_7+C_{10}+2C_{11})\,,\nn \\
\Gamma^{(1)}_4 &=& (2- 4N) B_4-2(N+1)(B_3+2B_2) 
+(N-2)(C_1+2C_2)-N(2C_4+C_7)\,.
\eeqa
We therefore obtain
\beqa
\Gamma^{(1)}_{1-2-3+4} &=& 0\,. 
\eeqa
We have 3 more structures:  $B^{ffg}_{\hat\alpha[\beta\alpha]}B^{ggf}_{\alpha[\beta\alpha]}$,
$B^{ffg}_{\hat\alpha[\hat\beta\hat\alpha]}B^{ggf}_{\alpha[\hat\beta\hat\alpha]}$ and
$B^{ffg}_{\alpha[\hat\beta\hat\alpha]}B^{ggf}_{\hat\alpha[\hat\beta\hat\alpha]}$ . We introduce operators $B_i^\prime = - B_i(\alpha\leftrightarrow \beta, f\leftrightarrow g)$ for the first class, $\tilde B_i = B_i(\alpha\leftrightarrow\hat\alpha)$ for the second, and $\hat B_i = B^\prime(\alpha\leftrightarrow\hat\alpha)$ for the 3rd. We then have
\beqa
\left(\Gamma_{III}^{10}\right)^{(1)}&=& 0\,.
\eeqa

\noindent
{\bf 4.} $B_{IV}^{10}=B^{ffg}_{\alpha[\hat\beta\hat\alpha]}B^{ggf}_{\alpha[\hat\beta\hat\alpha]}
+B^{ffg}_{\hat\alpha[\beta\alpha]}B^{ggf}_{\hat\alpha[\beta\alpha]} $\\
 \\ \noindent
We prepare the following 16 operators:
\beqa
B_1&=&B^{ffg}_{\alpha\hat\beta\hat\alpha}B^{ggf}_{\alpha\hat\beta\hat\alpha}, \
B_2=B^{ffg}_{\alpha\hat\beta\hat\alpha}B^{ggf}_{\alpha\hat\alpha\hat\beta}, \
B_3=B^{ffg}_{\alpha\hat\alpha\hat\beta}B^{ggf}_{\alpha\hat\beta\hat\alpha}, \
B_4=B^{ffg}_{\alpha\hat\alpha\hat\beta}B^{ggf}_{\alpha\hat\alpha\hat\beta}, \
C_1=B^{ffg}_{\alpha\hat\alpha\hat\alpha}B^{ggf}_{\alpha\hat\beta\hat\beta}, \nn \\
C_2&=&B^{ffg}_{\alpha\hat\beta\hat\beta}B^{ggf}_{\alpha\hat\alpha\hat\alpha}, \
C_3=B^{ggf}_{\hat\alpha\hat\beta\alpha}B^{ffg}_{\hat\alpha\hat\beta\alpha}, \
C_4=B^{fff}_{\alpha\hat\alpha\hat\beta}B^{ggg}_{\alpha\hat\alpha\hat\beta}, \
C_5=B^{ggf}_{\hat\alpha\hat\alpha\alpha}B^{ffg}_{\hat\beta\hat\beta\alpha}, \
C_6=B^{fff}_{\alpha\hat\beta\hat\beta}B^{ggg}_{\alpha\hat\alpha\hat\alpha}, \nn \\
C_7&=&B^{ggf}_{\hat\beta\hat\beta\alpha}B^{ffg}_{\hat\alpha\hat\alpha\alpha}, \
C_8=B^{fff}_{\alpha\hat\alpha\hat\alpha}B^{ggg}_{\alpha\hat\beta\hat\beta}, \
D_1=B^{ffg}_{\alpha\hat\beta\alpha}B^{ggf}_{\hat\alpha\hat\beta\hat\alpha}, \
D_2=B^{ggf}_{\alpha\hat\alpha\alpha}B^{ffg}_{\hat\alpha\hat\beta\hat\beta}, \
D_3=B^{ffg}_{\hat\alpha\hat\beta\hat\alpha}B^{ggf}_{\alpha\hat\beta\alpha}, \nn \\
D_4&=&B^{ggf}_{\hat\alpha\hat\beta\hat\beta}B^{ffg}_{\alpha\hat\alpha\alpha}, \nn 
\eeqa
in terms of which we write
\beqa
B^{ffg}_{\alpha[\hat\beta\hat\alpha]}B^{ggf}_{\alpha[\hat\beta\hat\alpha]}
&=&B_1-B_2-B_3+B_4\,.
\eeqa
Constraints give
\beqa
C_1+C_2+C_3+C_4= -(B_1+B_4),\ C_5+C_6+C_7+C_8= -4(B_2+B_3).
\eeqa
The 1-loop corrections can be calculated as
\beqa
\Gamma_1^{(1)} &=& (1-2N)B_1-2(N+1)(B_2+B_3) 
+N(B_4-C_3-C_4)+(N-2)(C_1+C_2)\,,\nn \\
\Gamma_2^{(1)} &=& (1+N)(B_2-2B_1-2B_4) 
+N(C_5+C_6-2C_3-2C_4)-2(C_1+C_2)\,,\nn \\
\Gamma_3^{(1)} &=& (1+N)(B_3-2B_1-2B_4) 
+N(C_7+C_8-2C_3-2C_4)-2(C_1+C_2)\,,\nn \\
\Gamma_4^{(1)} &=& (1-2N)B_4-2(N+1)(B_2+B_3) 
+N(B_1-C_3-C_4)+(N-2)(C_1+C_2)\,,\nn\\
&&
\eeqa
which give
\beqa
\Gamma^{(1)}_{1-2-3+4}&=& 8\frac{d}{\epsilon}B_{1-2-3+4}\,.
\eeqa
We introduce $\tilde B_i=B_i(\alpha\leftrightarrow\hat\alpha)$ for 
$B^{ffg}_{\hat\alpha[\beta\alpha]}B^{ggf}_{\hat\alpha[\beta\alpha]}$, so that
\beqa
\left(\Gamma_{IV}^{10}\right)^{(1)}&=&
8\frac{d}{\epsilon} (B+\tilde B)_{1-2-3+4}\,.
\eeqa

\noindent
{\bf 5.} $B_V^{10}=B^{ffg}_{\alpha[\hat\beta\hat\alpha]}B^{ggf}_{\hat\alpha[\beta\alpha]} 
+B^{ffg}_{\hat\alpha[\beta\alpha]}B^{ggf}_{\alpha[\hat\beta\hat\alpha]}$ and 
$B_{VI}^{10}=B^{ffg}_{\alpha[\beta\alpha]}B^{ggf}_{\hat\alpha[\hat\beta\hat\alpha]}
+ B^{ffg}_{\hat\alpha[\hat\beta\hat\alpha]}B^{ggf}_{\alpha[\beta\alpha]}$ \\
 \\ \noindent
We prepare the following 30 operators:
\beqa
B_1&=& B^{ffg}_{\alpha\hat\beta\hat\alpha}B^{ggf}_{\hat\alpha\beta\alpha},\
B_2= B^{ffg}_{\alpha\hat\beta\hat\alpha}B^{ggf}_{\hat\alpha\alpha\beta},\
B_3= B^{ffg}_{\alpha\hat\alpha\hat\beta}B^{ggf}_{\hat\alpha\beta\alpha},\
B_4= B^{ffg}_{\alpha\hat\alpha\hat\beta}B^{ggf}_{\hat\alpha\alpha\beta},\
B_5= B^{ffg}_{\alpha\beta\alpha}B^{ggf}_{\hat\alpha\hat\beta\hat\alpha},\nn \\
B_6&=& B^{ffg}_{\alpha\beta\alpha}B^{ggf}_{\hat\alpha\hat\alpha\hat\beta},\
B_7= B^{ffg}_{\alpha\alpha\beta}B^{ggf}_{\hat\alpha\hat\beta\hat\alpha},\
B_8= B^{ffg}_{\alpha\alpha\beta}B^{ggf}_{\hat\alpha\hat\alpha\hat\beta},\
C_1= B^{ggf}_{\beta\hat\alpha\hat\beta}B^{ffg}_{\alpha\alpha\hat\alpha},\
C_2= B^{ggf}_{\alpha\hat\alpha\hat\beta}B^{ffg}_{\alpha\beta\hat\alpha},\nn \\
C_3&=& B^{ggf}_{\hat\alpha\hat\alpha\alpha}B^{ffg}_{\alpha\hat\beta\beta},\
C_4= B^{ggf}_{\hat\alpha\hat\beta\alpha}B^{ffg}_{\alpha\hat\alpha\beta},\
C_5= B^{ffg}_{\beta\hat\beta\hat\alpha}B^{ggf}_{\hat\alpha\alpha\alpha},\
C_6= B^{ffg}_{\alpha\hat\alpha\hat\alpha}B^{ggf}_{\hat\beta\beta\alpha},\
C_7= B^{ffg}_{\beta\hat\alpha\hat\beta}B^{ggf}_{\hat\alpha\alpha\alpha},\nn \\
C_8&=& B^{ffg}_{\alpha\hat\alpha\hat\alpha}B^{ggf}_{\hat\beta\alpha\beta},\
C_9= B^{ggf}_{\alpha\hat\beta\hat\alpha}B^{ffg}_{\alpha\beta\hat\alpha},\
C_{10}= B^{ggf}_{\hat\alpha\hat\beta\alpha}B^{ffg}_{\beta\hat\alpha\alpha},\
C_{11}= B^{ggf}_{\hat\alpha\hat\alpha\alpha}B^{ffg}_{\beta\hat\beta\alpha},\
C_{12}= B^{ggf}_{\beta\hat\beta\hat\alpha}B^{ffg}_{\alpha\alpha\hat\alpha},\nn \\
D_1&=& B^{fff}_{\alpha\alpha\hat\beta}B^{ggg}_{\beta\hat\alpha\hat\alpha},\
D_2= B^{ggf}_{\beta\hat\alpha\alpha}B^{ffg}_{\alpha\hat\beta\hat\alpha},\
D_3= B^{ffg}_{\alpha\beta\hat\beta}B^{ggf}_{\alpha\hat\alpha\hat\alpha},\
D_4= B^{ffg}_{\alpha\hat\alpha\alpha}B^{ggf}_{\hat\alpha\hat\beta\beta},\
D_5= B^{ggf}_{\alpha\hat\beta\alpha}B^{ffg}_{\beta\hat\alpha\hat\alpha},\nn \\
D_6&=& B^{fff}_{\alpha\beta\hat\alpha}B^{ggg}_{\alpha\hat\alpha\hat\beta},\
D_7= B^{ffg}_{\alpha\hat\beta\alpha}B^{ggf}_{\hat\alpha\hat\alpha\beta},\
D_8= B^{fff}_{\alpha\beta\hat\beta}B^{ggg}_{\alpha\hat\alpha\hat\alpha},\
D_9= B^{ffg}_{\alpha\alpha\hat\beta}B^{ggf}_{\beta\hat\alpha\hat\alpha},\
D_{10}= B^{fff}_{\alpha\alpha\hat\alpha}B^{ggg}_{\beta\hat\alpha\hat\beta},\nn
\eeqa
in terms of which we have
\beqa
&&B^{ffg}_{\alpha[\hat\beta\hat\alpha]}B^{ggf}_{\hat\alpha[\beta\alpha]} +
B^{ffg}_{\alpha[\beta\alpha]}B^{ggf}_{\hat\alpha[\hat\beta\hat\alpha]} \equiv
{\cal B}_5 +{\cal B}_6\,,\\
&&{\cal B}_5 =B_{1-2-3+4}\,,\quad {\cal B}_6=B_{5-6-7+8}\,.
\eeqa
Constraints give
\beqa
C_1&=&C_3 = B_8-2B_1,\ D_1=B_8-4B_1,\ D_2 = -B_1,\
C_7+C_8+C_9+C_{10} = 2B_5-2B_4, \nn \\
&&C_2+C_5+C_{11} = 2B_6-3B_2,\ C_4+C_6+C_{12} = 2B_7-3B_3.
\eeqa
The 1-loop corrections can be calculated as
\beqa
\Gamma^{(1)}_1 &=& 2 B_1-2(N+1)(B_2+B_3)-2(C_5+C_6) + N(C_1+C_3-2C_2-2C_4)\,, \nn \\
\Gamma^{(1)}_2 &=& (2-N) B_2-2(N+1)(B_1+B_4)+(N-2)C_5 -2C_8 
+ N(-C_2+C_{11}-2C_{10})\,, \nn \\
\Gamma^{(1)}_3 &=& (2-N) B_3-2(N+1)(B_1+B_4)+(N-2)C_6 -2C_7 
+ N(-C_4+C_{12}-2C_{9})\,, \nn \\
\Gamma^{(1)}_4 &=& (2-2N)B_4-2(N+1)(B_2+B_3)+(N-2)(C_7+C_8) - N(C_9+C_{10})\,, 
\eeqa
which give
\beqa
\Gamma^{(1)}_{1-2-3+4} &=& 6\frac{d}{\epsilon}{\cal B}_5 + 
6\frac{d}{\epsilon}{\cal B}_6\,.
\eeqa
We introduce operators 
$\tilde B_i = B_i(\alpha\leftrightarrow\hat\alpha)$ for ${\cal B}_{5,6}$, 
so that we have
\beqa
\left(\Gamma_{V}^{10}\right)^{(1)}&=&6\frac{d}{\epsilon} ({\cal B} 
+\tilde{\cal  B})_5
+6\frac{d}{\epsilon}({\cal B} +\tilde{\cal  B})_6\,, \qquad
\left(\Gamma_{VI}^{10}\right)^{(1)}=
24\frac{d}{\epsilon}({\cal B} +\tilde{\cal  B})_6\,.
\eeqa 

\begin{table}[h] 
\centering 
\begin{tabular}[t]{c|c|c} 
\hline 
Dirac indices&$\gamma_j/(2d)$&$I$\\[1.0ex] 
\hline  
$111111$&$-24$&$0$\\[1.0ex] 
\hline  
$111112$&$-24$&$0,1$\\[1.0ex] 
\hline  
$111122$&$-4$&$0$\\[1.0ex] 
$$&$-24$&$0,1$\\[1.0ex] 
$$&$-40$&$2$\\[1.0ex] 
\hline  
$111222$&$-4$&0$$\\[1.0ex] 
$$&$-12$&$1$\\[1.0ex] 
$$&$-24$&$0,1$\\[1.0ex] 
$$&$-40$&$2$\\[1.0ex] 
$$&$-72$&$3$\\[1.0ex] 
\hline  
$111113$&$-16$&$0,1$\\[1.0ex] 
\hline  
$111123$&$-6$&$0,1$\\[1.0ex] 
$$&$-16$&$0,1$\\[1.0ex] 
$$&$-24$&$1,2$\\[1.0ex] 
\hline  
\end{tabular} 
\caption{\footnotesize Eigenvalues $\gamma_j$ of the anomalous
dimension matrix $\gamma$ and isospins of the corresponding eigenvectors 
for the case 3f3g.}  
\label{T3f3ga} 
\end{table}

\begin{table}[h] 
\centering 
\begin{tabular}[t]{c|c|c} 
\hline 
Dirac indices&$\gamma_j/(2d)$&$I$\\[1.0ex] 
\hline  
$111223 $&$0$&$0,1$\\[1.0ex] 
$$&$-6$&$0,1$\\[1.0ex] 
$$&$-16$&$0,1$\\[1.0ex] 
$$&$-18$&$1,2$\\[1.0ex] 
$$&$-24$&$1,2$\\[1.0ex] 
$$&$-48$&$2,3$\\[1.0ex] 
\hline  
$111133$&$-4$&$0$\\[1.0ex] 
$$&$-16$&$0,1,2$\\[1.0ex] 
\hline  
$111134$&$0$&1$$\\[1.0ex] 
$$&$-4$&$0$\\[1.0ex] 
$$&$-12$&$1$\\[1.0ex] 
$$&$-16$&$0,1,2$\\[1.0ex] 
\hline  
$111233$&$4$&$1$\\[1.0ex] 
$$&$-4$&$0$\\[1.0ex] 
$$&$-8$&$0,1,1$\\[1.0ex] 
$$&$-16$&$0,1,2$\\[1.0ex] 
$$&$-32$&$1,2,3$\\[1.0ex] 
\hline  
$111234$&$20$&$0$\\[1.0ex] 
$$&$8$&$1$\\[1.0ex] 
$$&$4$&$1$\\[1.0ex] 
$$&$0$&$1$\\[1.0ex] 
$$&$-4$&$0$\\[1.0ex] 
$$&$-8$&$0,1,1,2$\\[1.0ex] 
$$&$-12$&$1$\\[1.0ex] 
$$&$-16$&$0,0,0,1,1,2,2,2$\\[1.0ex] 
$$&$-32$&$1,2,3$\\[1.0ex] 
\hline  
$112233$&$8$&$0$\\[1.0ex] 
$$&$4$&$1$\\[1.0ex] 
$$&$-4$&$0,0,1,2$\\[1.0ex] 
$$&$-8$&$0,1,1,2$\\[1.0ex] 
$$&$-16$&$0,1,2$\\[1.0ex] 
$$&$-28$&$2$\\[1.0ex] 
$$&$-30$&$1,2,3$\\[1.0ex] 
\hline  
\end{tabular} 
\caption{\footnotesize As in Table~2 (continued).}  
\label{T3f3gb} 
\end{table}

\begin{table}[h] 
\centering 
\begin{tabular}[t]{c|c|c} 
\hline 
Dirac indices&$\gamma_j/(2d)$&$I$\\[1.0ex] 
\hline  
$112234 $&$20$&$0$\\[1.0ex] 
$$&$12$&$1$\\[1.0ex] 
$$&$8$&$0,1$\\[1.0ex] 
$$&$4$&$1$\\[1.0ex] 
$$&$0$&$1,1$\\[1.0ex] 
$$&$-4$&$0,0,1,2$\\[1.0ex] 
$$&$-8$&$0,1,1,2$\\[1.0ex] 
$$&$-12$&$1$\\[1.0ex] 
$$&$-16$&$0,0,1,1,2,2,2$\\[1.0ex] 
$$&$-28$&$2$\\[1.0ex] 
$$&$-32$&$1,2,3$\\[1.0ex] 
$$&$-36$&$1,2,3$\\[1.0ex] 
\hline  
$111333$&$-6$&$0,1$\\[1.0ex] 
$$&$-24$&$0,1,2,3$\\[1.0ex] 
\hline  
$111334$&$0$&$0,1,1,2$\\[1.0ex] 
$$&$-6$&$0,1$\\[1.0ex] 
$$&$-18$&$1,2$\\[1.0ex] 
$$&$-24$&$0,1,2,3$\\[1.0ex] 
\hline  
$112334$&$24$&$0,1$\\[1.0ex] 
$$&$6$&$0,1$\\[1.0ex] 
$$&$0$&$0,0,1,1,1,1,2,2$\\[1.0ex] 
$$&$-6$&$0,1$\\[1.0ex] 
$$&$-12$&$1,1,2,2$\\[1.0ex] 
$$&$-18$&$1,1,2,2$\\[1.0ex] 
$$&$-24$&$0,1,2,3$\\[1.0ex] 
$$&$-30$&$0,1,2,3$\\[1.0ex] 
\hline 
\end{tabular} 
\caption{\footnotesize As in Table~2 (continued).}  
\label{T3f3gc} 
\end{table}

\clearpage

\section{Some useful formulae for angular momentum states}
\label{appendixB}
\subsection{Eigenstates}
At given $J$, there are 2 distinct states, the spin-singlet ($S=0$)
state and the spin-triplet ($S=1$) state.

The singlet state is denoted as $^1J_J$, since it has $S=0$ and $J=L$.
The fact that $I+L+S$ must be odd to satisfy fermion anti-symmetry
gives $I=0$ for odd $J$ and $I=1$ for even $J$. The eigenstate with
$J_z$ can be easily obtained as
\beqa
\vert ^1J_J, J_z \rangle & =& \vert J_z,0\rangle_{J,S=0}\,,
\eeqa
where $\vert J_z, S_z\rangle_{J,S} = \vert J_z\rangle_J \otimes \vert
S_z \rangle_S$. 

The spin-triplet state is classified into 3 types:  $^3J_J$, 
$^3(J\pm1)_J$. For the first one, $I=0$ (even $J$) or $I=1$ (odd $J$), 
and vice versa for the other two types. 
By the Wigner-Eckart theorem,
the matrix elements of the five operators do not depend on
$J_z$. Therefore it is enough to know eigenstates with $J_z=J$
only. Explicitly we have
\begin{eqnarray}
\vert ^3J_J, J\rangle &=& \frac{1}{\sqrt{J+1}}\left\{ \vert
J-1,1\rangle_{J,1} 
- \sqrt{J}\vert J,0\rangle_{J,1}\right\} \ \ ,\\
\vert ^3(J-1)_J, J\rangle &=&  \vert J-1,1\rangle_{J-1,1} \ \ , \\
\vert ^3(J+1)_J, J\rangle &=& \frac{1}{\sqrt{(J+1)(2J+3)}} \Bigl\{
\vert J-1,1\rangle_{J+1,1} 
\nonumber \\
&+&\sqrt{2J+1}\left[ \sqrt{(J+1)}\vert J+1,-1\rangle_{J+1,1}-\vert
J,0\rangle_{J+1,1}\right]\Bigr\}\,.
\end{eqnarray}

\subsection{Evaluation of each operator}
Using these eigenstates, it is easy to see
\begin{eqnarray}
{\vec \sigma_1}\cdot{\vec \sigma_2} &=& 2S(S+1)-3 = -3,\ 1,\ 1,\ 1 \ \ \
\ ,\\
{\vec L}\cdot {\vec S} &=& \frac{J(J+1)-L(L+1)-S(S+1)}{2} = 0,\ -1,\
J-1,\ -(J+2) \ \ \ \ , 
\end{eqnarray}
for $^1J_J$, $^3J_J$, $^3(J-1)_J$ and $^3(J+1)_J$, respectively.

For $S_{12}$ defined in (\ref{S_12})
the results are more complicated due to the mixing between
$^3(J-1)_J$ and $^3(J+1)_J$. After a little algebra we obtain,
\begin{eqnarray}
S_{12} &=& 0, \ 2, \ \left( \begin{array}{cc}
 -\dfrac{2(J-1)}{2J+1}, & \dfrac{6\sqrt{J(J+1)}}{2J+1}  \\
 \\
 \dfrac{6\sqrt{J(J+1)}}{2J+1}, & -\dfrac{2(J+2)}{2J+1} \\
\end{array}\right)\,.
\end{eqnarray}

\section{The $I=2$ 2-pion system}
\label{appendixC}

Here we consider the operator product expansion of two 
iso-vector pseudoscalar densities in QCD.

\subsection{Anomalous dimensions}


The local composite operators with $\pi^+\pi^+$ quantum numbers in QCD
with lowest dimension are 4--quark operators with dimension 6. 
There are 5 independent such (bare) scalar operators
\ba
{\cal O}_1&=&\bar{d}\gamma^\mu u\cdot \bar{d}\gamma_\mu u
+\bar{d}\gamma^\mu\gamma_5 u\cdot \bar{d}\gamma_\mu\gamma_5 u,
\\
{\cal O}_2&=&\bar{d} u\cdot \bar{d} u
-\bar{d}\gamma_5 u\cdot \bar{d}\gamma_5 u,
\\
{\cal O}_3&=&\bar{d}\gamma^\mu u\cdot \bar{d}\gamma_\mu u
-\bar{d}\gamma^\mu\gamma_5 u\cdot \bar{d}\gamma_\mu\gamma_5 u,
\\
{\cal O}_4&=&\bar{d} u\cdot \bar{d} u
+\bar{d}\gamma_5 u\cdot \bar{d}\gamma_5 u,
\\
{\cal O}_5&=&\bar{d}\sigma^{\mu\nu} u\cdot \bar{d}\sigma_{\mu\nu} u,
\label{pipiscalar}
\ea
where in this appendix we use the notation $u=q^u$, $d=q^d$ and suppress
explicit color and Dirac indices of the quark fields. 

There are also 3 independent such (bare) traceless tensors operators
\ba
T_1^{\mu\nu}&=&\bar{d}\gamma^\mu u\cdot \bar{d}\gamma^\nu u
+\bar{d}\gamma^\mu\gamma_5 u\cdot \bar{d}\gamma^\nu\gamma_5 u
-\frac{1}{D}\,g^{\mu\nu}{\cal O}_1,
\\
T_2^{\mu\nu}&=&\bar{d}\gamma^\mu u\cdot \bar{d}\gamma^\nu u
-\bar{d}\gamma^\mu\gamma_5 u\cdot \bar{d}\gamma^\nu\gamma_5 u
-\frac{1}{D}\,g^{\mu\nu}{\cal O}_3,
\\
T_3^{\mu\nu}&=&\bar{d}\sigma^{\mu\tau} u\cdot \bar{d}\sigma^\nu_{\,\,\,\tau} u
-\frac{1}{D}\,g^{\mu\nu}{\cal O}_5.
\label{pipitensor}
\ea
As in the main text, operators are renormalized according to the
formula
\be
{\cal O}_A^{(R)}={\cal O}_A-\frac{g^2}{32\pi^2\epsilon}\gamma_{AB}{\cal O}_B+
\dots,
\end{equation}
and similarly for tensor fields. The results for the one-loop anomalous
dimensions of the scalar fields can be found in \cite{BMU}. The non-vanishing
entries of the mixing matrix for the scalar case are
\ba
\gamma_{11}&=&-4\,,\\
\gamma_{22}&=&16\,,\\
\gamma_{33}&=&-2\,, \qquad\quad\ \ \ \ \gamma_{32}=-12\,,\\
\gamma_{44}&=&10\,,\qquad\quad \,\gamma_{45}=1/3\,,\\
\gamma_{54}&=&-20\,,\qquad\quad\ \, \ \gamma_{55}=-34/3\,.
\ea
We have extended the analysis of ref. \cite{BMU} to the tensor case. Here
we find
\ba
\gamma^{(T)}_{11}&=&-8/3\,,\\
\gamma^{(T)}_{22}&=&2/3\,,\qquad\quad \gamma^{(T)}_{23}=-2\,,\\
\gamma^{(T)}_{33}&=&-16/3\,.
\ea
It is useful to introduce the (one-loop) diagonally renormalized operators.
For the scalar case we have
\be
X_A^{(R)}=X_A-\frac{g^2}{32\pi^2\epsilon}\hat\gamma_AX_A+\dots,
\end{equation}
where
\ba
X_1&=&{\cal O}_1\,,\qquad\quad\qquad\quad\,\hat\gamma_1=-4\,,\\
X_2&=&{\cal O}_2\,,\qquad\quad\qquad\quad\,\hat\gamma_2=16\,,\\
X_3&=&{\cal O}_3+\frac{2}{3}{\cal O}_2\,,\qquad\quad\hat\gamma_3=-2\,,\\
X_4&=&{\cal O}_4+B{\cal O}_5\,,\qquad\quad\hat\gamma_4=(\sqrt{964}-2)/3\,,\\
X_5&=&{\cal O}_5+C{\cal O}_4\,,\qquad\quad\hat\gamma_5=-(\sqrt{964}+2)/3\,.
\ea
Here
\be
B=\frac{16-\sqrt{241}}{30}=0.01586\,,\qquad\qquad
C=32-\sqrt{964}=0.95165\,.
\end{equation}
                      
\subsection{OPE for $I=2$ $\pi$-$\pi$ scattering}

\noindent
The OPE for two $\pi$ fields can be written in QCD as

\begin{equation}
\pi(x)\pi(0)=\sum_\alpha \gamma_\alpha(x){\cal B}_\alpha+\cdots
\label{OPEpipi}
\end{equation}
Here $\pi(x)$ is the field annihilating $\pi^+$, ${\cal B}_\alpha$ are 
the renormalized doubly charged dimension 6 operators discussed in the 
previous section, $x$ is spacelike (and for simplicity we assume its
time component vanishes) and $\gamma_\alpha(x)$ are c-number coefficient
functions. [Here we \lq\lq pretend" all operators are scalar, although
in fact three of them are symmetric traceless tensors. Taking into account
their tensor structure however does not change any of our conclusions here.]
The dots stand for higher dimensional operators with less singular
coefficient functions.

The short distance asymptotics of the $I=2$ wave function is given by
\begin{equation}
\Psi(x)=\langle0\vert\pi(x)\pi(0)\vert2\rangle \sim \sum_\alpha\gamma_\alpha
(x)B_\alpha+\cdots
\end{equation}
where
\begin{equation}
B_\alpha=\langle0\vert{\cal B}_\alpha\vert2\rangle
\end{equation}
are the (energy dependent) matrix elements of the local operators.

In QCD we can write the divergence of the axial current as
\begin{equation}
\partial^\mu A_\mu=\partial^\mu\left(\bar d\gamma_\mu\gamma_5 u\right)=
m_0\Phi_0=m_{\rm R}\Phi_R,
\end{equation}
where
\begin{equation}
\Phi_0(x)=\bar d(x)\gamma_5 u(x)
\end{equation}
is a (bare) quark bilinear field with $\pi^+$ quantum numbers
and the quark mass parameter $m_0$ is the sum of the u and d
quark masses. [Here we used the fact that the axial current, being
partially conserved, has renormalization constant $Z=1$. There is
a subtlety in dimensional regularization where because of the presence
of $\gamma_5$ the renormalization constant is not equal to unity. It
is finite nevertheless and this does not alter our conclusions at the
1-loop level.]
The (canonically normalized) pion field is defined as
\begin{equation}
\pi(x)=\frac{1}{m_\pi^2f_\pi}\partial_\mu A_\mu(x)=\Omega\Phi_R(x)\,,
\end{equation}
where $\Omega$ is a constant:
\begin{equation}
\Omega=\frac{m_{\rm R}}{m_\pi^2f_\pi}\,.
\end{equation}
From this we see that the field $\Phi_R$ renormalizes with the inverse of
the mass renormalization constant.

The RG analysis of the pion-pion wave function goes along the same lines
as in the main text for the nucleon-nucleon case.
By inspecting the spectrum of anomalous dimensions we see that, again, only 
operators already present in the tree level expansion
\be
(\bar{d}\gamma_5 u)^2=-\frac{1}{2}{\cal O}_2+\frac{1}{2}{\cal O}_4=
-\frac{1}{2}X_2+\frac{1}{2(1-BC)}[X_4-BX_5]
\end{equation}
contribute to the leading short distance part of the wave function
and even from this set we need only the operators with the largest anomalous 
dimensions. Since the coefficient of such an operator 
$X_A^{(R)}$ is asymptotically proportional to
\be
(-\ln r)^{\frac{1}{2b_0}(\hat\gamma_A+d_0)},
\end{equation}
where $b_0=11-2N_f/3$ and $d_0=-16$ comes from the mass renormalization.

Numerically the spectrum of $\hat\gamma_A$s is
\begin{equation}
\langle -4;16;-2;9.68;-11.02;-\frac{16}{3};-\frac{8}{3};\frac{2}{3}\rangle\,,
\end{equation}
corresponding to the spectrum of powers
\begin{equation}
\langle-1.11;0;-1;-0.35;-1.50;-1.19;-1.04;-0.85\rangle
\end{equation}
numerically. (Here we took $\Nf=3$ for simplicity.)

We have, again, a leading zero eigenvalue and all the other powers are
subleading. 
The next one is $-0.35$ so the wave function is asymptotically
\begin{equation}
\Psi(x)\sim\psi_0+\psi_1\ell^{-b}+\cdots,
\end{equation}
where $b=0.35$. This corresponds to
\begin{equation}
V(r)\sim \frac{\psi_1}{\psi_0}\frac{b}{r^2\ell^{(1+b)}}\,.
\end{equation}
Here
\be
\psi_0=-\frac{\Omega^2}{2}\langle0\vert{\cal O}_2^{(R)}\vert2\rangle
\end{equation}
and $\psi_1$ is proportional to (with a positive coefficient)
the linear combination
\be
\langle0\vert{\cal O}_4^{(R)}\vert2\rangle
+B\langle0\vert{\cal O}_5^{(R)}\vert2\rangle.
\end{equation}
Note that the ratio $\psi_1/\psi_0$ may be energy dependent. 
We need to calculate this ratio (or at least its sign)
nonperturbatively, to be able to determine whether the potential in this
channel is attractive or repulsive. 
ChPT is not applicable for this problem since there are too many extra 
low energy constants characterizing the matrix elements of 4--quark operators
and in the end the sign of this ratio is left undetermined.
In the absence of a reliable
non-perturbative method to calculate the above matrix elements we try to
estimate them by inserting a complete set of states in the middle of the
operator and truncating the sum after the 1-particle contribution. This
is very similar in spirit to the vacuum insertion method 
\cite{VacIns}, (oft rightly criticized) however surprisingly successfully 
applied to $\Delta S=2$ weak matrix elements in the past.
In this approximation
\be
\langle0\vert (\bar{d}\Gamma_1 u)\cdot(\bar{d}\Gamma_2 u)\vert2\rangle
\approx
\langle0\vert (\bar{d}\Gamma_1 u)\vert1\rangle
\langle1\vert(\bar{d}\Gamma_2 u)\vert2\rangle
\end{equation}
and therefore we have (in this approximation)
\be
\langle0\vert{\cal O}_4^{(R)}\vert2\rangle\approx
-\langle0\vert{\cal O}_2^{(R)}\vert2\rangle
\end{equation}
and
\be
\langle0\vert{\cal O}_5^{(R)}\vert2\rangle\approx0.
\end{equation}
Thus the ratio $\psi_1/\psi_0$ is positive in this naive approximation
and the potential is repulsive, as indicated by the 
(quenched) lattice measurements \cite{SI}.


\end{document}